\documentclass[aps,prd,superscriptaddress,preprintnumbers,tightenlines,nofootinbib, twocolumn,10pt,floatfix,amsmath,amssymb,longbibliography,superscriptaddress,nofootinbib]{revtex4-2}

\usepackage{graphicx}
\usepackage{amsmath}
\usepackage{amsfonts}
\usepackage{amssymb}
\usepackage{bm}
\usepackage{mathrsfs}
\usepackage{graphicx}
\usepackage[colorlinks]{hyperref}
\usepackage{empheq}
\usepackage{ulem}
\usepackage[usenames]{color}
\usepackage{subfigure}
\usepackage{orcidlink}

\allowdisplaybreaks

\DeclareSymbolFontAlphabet{\mathrsfs}{rsfs}
\DeclareMathAlphabet{\mathcal}{OMS}{cmsy}{m}{n}

\newcommand{\FP}{\mathop{\mathrm{FP}}_{B=0}}

\newcommand{\nn}{\nonumber}
\newcommand{\calO}{\mathcal{O}}
\newcommand{\dd}{\mathrm{d}}
\newcommand{\di}{\mathrm{i}} 
\newcommand{\de}{\mathrm{e}} 
\newcommand{\dM}{\mathrm{M}}
\newcommand{\dS}{\mathrm{S}}
\newcommand{\dI}{\mathrm{I}}
\newcommand{\dJ}{\mathrm{J}}
\newcommand{\dU}{\mathrm{U}}
\newcommand{\dV}{\mathrm{V}}
\newcommand{\mtot}{m_{\mathrm{tot}}}

\begin{document}

\title{\texorpdfstring{Comparison of 4.5PN and 2SF gravitational energy fluxes \\from quasicircular compact binaries}{Comparison of 4.5PN and 2SF gravitational energy fluxes from quasicircular compact binaries}}

\author{Niels \textsc{Warburton}\,\orcidlink{0000-0003-0914-8645}}
\affiliation{School of Mathematics and Statistics, University College Dublin, \\Belfield, Dublin 4, Ireland, D04 V1W8}

\author{Barry \textsc{Wardell}\,\orcidlink{0000-0001-6176-9006}}
\affiliation{School of Mathematics and Statistics, University College Dublin, \\Belfield, Dublin 4, Ireland, D04 V1W8}

\author{David \textsc{Trestini}\,\orcidlink{0000-0002-4140-0591}}
\affiliation{CEICO, Institute of Physics of the Czech Academy of Sciences, \\Na Slovance 2, 182 21 Praha 8, Czechia}

\author{Quentin \textsc{Henry}\,\orcidlink{0000-0003-4071-2873}}
\affiliation{Max Planck Institute for Gravitational Physics (Albert Einstein Institute),\\ Am M\"uhlenberg 1, Potsdam 14476, Germany}

\author{Adam \textsc{Pound}\,\orcidlink{0000-0001-9446-0638}}
\affiliation{School of Mathematical Sciences and STAG Research Centre,
University of Southampton, Southampton, United Kingdom, SO17 1BJ}

\author{Luc \textsc{Blanchet}\,\orcidlink{0000-0003-1142-9534}}
\affiliation{$\mathcal{G}\mathbb{R}\varepsilon{\mathbb{C}}\mathcal{O}$, 
Institut d'Astrophysique de Paris, UMR 7095, CNRS, \\Sorbonne Universit{\'e}, 98\textsuperscript{bis} boulevard Arago, 75014 Paris, France}

\author{Leanne \textsc{Durkan}\,\orcidlink{0000-0001-8593-5793}}
\affiliation{Center for Gravitational Physics, Department of Physics, University of Texas at Austin, Austin, TX, USA, 78712}

\author{Guillaume \textsc{Faye}\,\orcidlink{000-0003-2921-1525}}
\affiliation{$\mathcal{G}\mathbb{R}\varepsilon{\mathbb{C}}\mathcal{O}$, 
Institut d'Astrophysique de Paris, UMR 7095, CNRS, \\Sorbonne Universit{\'e}, 98\textsuperscript{bis} boulevard Arago, 75014 Paris, France}

\author{Jeremy \textsc{Miller}\,\orcidlink{0000-0002-6862-1335}}
\affiliation{Shamoon College of Engineering, Be'er Sheva, Israel}

\date{\today}

\begin{abstract}
Recent years have seen significant advances in models of gravitational waveforms emitted by quasicircular compact binaries in two regimes: the weak-field, post-Newtonian regime, in which the gravitational wave energy flux has now been calculated to fourth-and-a-half post-Newtonian order (4.5PN) [Phys.~Rev.~Lett.~\textbf{131}, 121402 (2023)]; and the small-mass-ratio, gravitational self-force regime, in which the flux has now been calculated to second perturbative order in the mass ratio (2SF) [Phys.~Rev.~Lett.~\textbf{127}, 151102 (2021)]. We compare these results and find evidence of mutual consistency between the two (very distinct though both first-principle) perturbative calculations.
\end{abstract}

\maketitle

\section{Introduction}

Future gravitational-wave (GW) detectors will demand improved waveform models, both due to improved instrument sensitivity and due to a larger variety of systems that will be observed~\cite{Purrer:2019jcp,Dhani:2024jja}. This has motivated continual progress in modeling the waveform emission from binary systems of black holes and neutron stars~\cite{LISAConsortiumWaveformWorkingGroup:2023arg}.

Recently, two important milestones were achieved in this effort, specifically in perturbation methods for general relativity (GR) applied to gravitational waves from compact binary systems on quasicircular orbits: (i)~the gravitational self-force (GSF) approach was pushed to second order in the small mass ratio, both for the binary's conserved mass-energy~\cite{Poundetal20}
and the GW emission~\cite{WPWMD21,WPWMDT23}; (ii)~the post-Newtonian (PN) approximation was completed at 4PN order for the equations of motion~\cite{DJS14,DJS16,BBBFMc,MBBF17,FS19,FPRS19} and at 4.5PN order (beyond the Einstein quadrupole formula) for the energy flux~\cite{MHLMFB20,MBF16,LHBF22,LBHF22,TB23,BFHLT23a,BFHLT23b}, while the dominant $(\ell,m)=(2,2)$ mode of the waveform was obtained at 4PN order~\cite{BFHLT23a,BFHLT23b}. The aim of this paper is to compare the salient results of the two approaches, confirming the preliminary agreement found in~\cite{BFHLT23Moriond}.

For the purposes of the comparison, we focus on a binary system of two nonspinning black holes of masses $m_1$ and $m_2$ on a bound orbit. We assume that there is no incoming radiation and that the spacetime is asymptotically flat. By convention, we assume $m_1 \geqslant m_2$ (and $m_1 \gg m_2$ in the GSF case); we denote the total mass $\mtot = m_1+m_2$ and the symmetric mass ratio $\nu = m_1 m_ 2/\mtot^2$.
Far away from the binary, we assume a linearized metric around Minkowski, $g_{\alpha\beta} = \eta_{\alpha\beta} + h_{\alpha\beta}$, and  introduce a spherical coordinate system $(u,r,\theta,\phi)$ such that $u$ is a null coordinate (i.e., $g^{uu} = 0$). We will also use the usual spherical harmonics $Y^{\ell m}(\theta,\phi)$ as well as the spin-weighted ones $Y_{-2}^{\ell m}(\theta,\phi)$, where the integer $m$ should not be confused with a mass. We often set $G=c=1$ and pose $\calO(n)\equiv\calO(c^{-n})$ for small PN remainders.

We write the asymptotic gravitational waveform in a transverse traceless (TT) gauge, and decompose it into a ``plus'' mode $h_+$ and a ``cross'' mode $h_\times$. We can then expand the waveform $h \equiv h_+ - \di h_\times$ into spin-weighted spherical harmonics (with weight $-2$) as follows:
\begin{align}\label{eq:modeDecomposition}
h &= \frac{1}{r}\sum_{\ell=2}^{+\infty}\sum_{m=-\ell}^{\ell} h_{\ell m}(u)\, Y_{-2}^{\ell m}(\theta,\phi) + \calO\!\left(\frac{1}{r^2} \right)\,,
\end{align}
where $h$ is dimensionless and $h_{\ell m}$ has dimension of length since we have factored out the dependence on the radial distance in order to state purely asymptotic results. Each mode can be factorized  as 
\begin{equation}\label{eq:hlmAmplitudePhase}
    h_{\ell m}(u) = \mtot \,\hat{h}_{\ell m}(u) \,\de^{-\di m \psi(u)}\,,
\end{equation}
where the $\hat{h}_{\ell m}$ are dimensionless and we choose by convention that $\hat{h}_{22}$ is real-valued. Thus, the imaginary parts of the other $\hat{h}_{\ell m}$ modes account for higher-order dephasing between the different modes. We define the dimensionless parameter
\begin{equation}\label{eq:defx}
	x \equiv \bigl(\mtot \,\omega\bigr)^{2/3}\,,
\end{equation}
which is related to the frequency $\omega \equiv \dd \psi/\dd u$ associated to the $(2,2)$ mode and represents the small PN parameter $x=\calO(2)$. 

In the case of quasicircular orbits to which we will now specialize, the time dependence of $\hat{h}_{\ell m}$ is entirely captured by $x(u)$ and $\nu(u)$, namely we can write \mbox{$\hat{h}_{\ell m} \equiv \hat{h}_{\ell m}(x,\nu)$}. The frequency variable $x$ and masses~$m_i$ evolve secularly, following equations of the form 
\begin{subequations}\label{eq:xdotmdot}\begin{align}
    \frac{\dd x}{\dd u} &= \frac{\Xi(x,\nu)}{\mtot}\,,\label{eq:xdot}\\
    \frac{\dd m_i}{\dd u} &= {\cal F}_{\mathcal{H}_i}(x,\nu)\,,\label{eq:mdot}
\end{align}
\end{subequations}
where ${\cal F}_{\mathcal{H}_i}(x,\nu)$ is the flux of energy into the horizon of the black hole of mass $m_i$. Here  $\hat{h}_{\ell m}$, ${\cal F}_{\mathcal{H}_i}$, and the forcing function $\Xi$ are dimensionless and therefore can only be functions of dimensionless combinations of the binary's frequency $\omega$ and masses $m_i$ --- and all such combinations can be written in terms of $x$ and $\nu$. Equations~\eqref{eq:xdotmdot} allow us to apply the chain rule 
\begin{equation} \frac{\dd}{\dd u} = \frac{\Xi(x,\nu)}{\mtot} \, \frac{\partial}{\partial x} + \sum_{i=1,2}{\cal F}_{\mathcal{H}_i}(x,\nu)\, \frac{\partial}{\partial m_i}\,, 
\end{equation}
when acting on $h_{\ell m}$ with derivatives of $u$.

The asymptotic energy flux carried by GWs is
\begin{equation}\label{eq:defFlux}
\mathcal{F} = \frac{1}{16\pi} \int \dd \Omega_2\,|\dot{h}|^2 = \sum_{\ell=2}^{+\infty}\sum_{m=1}^{\ell} \mathcal{F}_{\ell m}(x,\nu) \,,
\end{equation}
where the dot stands for $\dd/\dd u$ and where we have defined 
\begin{equation}\label{eq:defFlm}
\mathcal{F}_{\ell m}(x,\nu) = \frac{1}{8\pi} |\dot{h}_{\ell m}|^2\,.
\end{equation}
Note that $m=0$ modes do not contribute for quasicircular orbits at the orders we consider, and a factor of 2 has appeared when expressing the sum in terms of positive $m$ modes, due to the relationship $h_{\ell (-m)}=(-)^m {h}_{\ell m}^*$ (the star $*$ indicates complex conjugation). Also note that although $\dot{h}_{\ell m}$ is a function of $(x,\nu,\psi)$, the modulus squared $|\dot{h}_{\ell m}|^2$ depends only on $(x,\nu)$; these fluxes are related in a one-to-one manner to the moduli of the $h_{\ell m}$ modes, but do not carry any information about the phases.

In the PN approach, which assumes small orbital frequencies, we obtain the fluxes as an expansion in the small parameter $x$ of Eq.~\eqref{eq:defx}, with coefficients that are functions of $\nu$:
\begin{equation}
    \mathcal{F}_{\ell m}(x,\nu) = \frac{32}{5}\,\nu^2x^5\biggl[1+\!\!\sum_{q\geqslant 2, \,k\geqslant 0} \!\!\mathcal{F}^{qk}_{\ell m}(\nu) \,x^{q/2}(\ln x)^k \biggr]\,.
\end{equation}
Conversely, the self-force approach assumes small mass ratios and obtains the flux as an expansion in $\nu$, with coefficients that are functions of $x$: 
\begin{equation}\label{eq:GSF flux generic}
    \mathcal{F}_{\ell m}(x,\nu) = \nu^2\biggl[\mathcal{F}^{(1)}_{\ell m}(x)+\nu \mathcal{F}^{(2)}_{\ell m}(x)+{\cal O}(\nu^2)\biggr]\,,
\end{equation}
where the SF order ``$(n)$'' indicates the order in $\nu$ at which the field equations are solved in order to obtain the corresponding flux contribution $\mathcal{F}^{(n)}_{\ell m}$. 

To compare the two expansions of the flux, we note that formally, one can perform a double expansion in $x$ and $\nu$, so as to obtain an expression of the form
\begin{equation}\label{eq:structureFlm}
    \mathcal{F}_{\ell m} = \frac{32}{5}\,\nu^2 x^5 \!\!\!\!\!\sum_{p\geqslant 0, \,q\geqslant 0, \,k\geqslant 0} \!\!\!\mathcal{F}^{pqk}_{\ell m} \nu^p x^{q/2} (\ln x)^k \,.
\end{equation}
As both the PN and GSF methods are first-principle perturbation methods of GR, they should yield strictly the same coefficients $\mathcal{F}^{pqk}_{\ell m}$ in the expansion (this statement still holds when applied to each individual $\mathcal{F}_{\ell m}$). The 4.5PN expression~\cite{BFHLT23a} and the 1SF results of analytical black hole perturbation~\cite{TSasa94,TTS96} theory have been shown to agree~\cite{BFHLT23a}, namely the coefficients $\mathcal{F}^{pqk}_{\ell m}$ for $p=0$ and $q\in [\![0,9]\!]$ are strictly identical. 
Reference~\cite{WPWMD21} additionally showed numerical agreement between the flux at 2SF and 3.5PN. 

The goal of this article is to confirm agreement up to 4.5PN. 
Our central conclusion is that the 2SF and 4.5PN results are consistent with each other at each available order: i.e. $\sum_{\ell m}\mathcal{F}^{pqk}_{\ell m}$ agrees between the two methods for $p=1$ and $q\in [\![0,9]\!]$ (and $k=0$ or $1$). However, at 4.5PN in particular, the comparison is hampered by poor overlap of the regions of validity of the two approximations: the 4.5PN flux is only in its asymptotic regime of validity for small values of $x$ in which numerical error in the 2SF data becomes significant enough that it interferes with the comparison.

\section{Review of 4.5PN flux}
\label{sec:PN}

The 4.5PN flux and waveform of compact binaries have been obtained by combining two approximation methods: the classic PN expansion which assumes small orbital velocities ($v/c\to 0$), equivalent to large separations between the two bodies, and the multipolar post-Minkowskian (MPM) method which combines the PM or non-linearity expansion ($G\to 0$) with multipolar series parametrized by specific multipole moments. The PN expansion is implemented in an inner domain (or near zone) covering the matter source but whose radius is much less than a gravitational wavelength. The MPM expansion is valid in an external domain which overlaps with the near zone of the source and extends into the far wave zone. The MPM field represents the most general solution of the Einstein field equation (say, in harmonic coordinates) in the exterior zone. The PN and MPM expansions are matched in the exterior part of the near zone, which is the region of common validity of both expansions: the whole procedure is called the post-Newtonian-multipolar-post-Minkowkian (PN-MPM) formalism~\cite{BD86,B87,BD88,BD92,B98mult}.

The MPM construction of the metric is carried out as a functional of a set of parameters, called the ``canonical'' mass and current type multipole moments $\dM_L(t)$ and $\dS_L(t)$, which are symmetric and trace-free (STF) with respect to their $\ell$ indices (where $L \equiv i_1\cdots i_\ell$). The canonical moments are defined from the linearized approximation $\frak{h}^{\alpha\beta}_1[\dM_L,\dS_L]$ of the MPM construction,
\begin{equation}
\frak{h}^{\alpha\beta}_\text{MPM} = \sum_{n=1}^{+\infty} G^n \,\frak{h}^{\alpha\beta}_n[\dM_L,\dS_L]\,,
\end{equation}
where $\frak{h}^{\alpha\beta}=\sqrt{-g}g^{\alpha\beta}-\eta^{\alpha\beta}$ is the ``gothic metric deviation'', and each PM approximation is computed iteratively~\cite{BD86}. The time dependence of the metric includes many non-local integrals over the moments $\dM_L$, $\dS_L$, from $t=-\infty$ up to the current time.
The matching to the PN field in the source's near zone determines the relations between $\dM_L$, $\dS_L$ and the actual ``source'' mass and current multipole moments $\dI_L$, $\dJ_L$, say
\begin{equation}\label{eq:cansource}
	\dM_L = \dI_L + \calO(5)\,,\qquad \dS_L = \dJ_L + \calO(5)\,,
\end{equation}
where the PN remainder terms are entirely controlled with the 4PN precision~\cite{BFL22, BFHLT23b}. Here $\dI_L$, $\dJ_L$ are functionals of the matter plus the gravitation  stress-energy pseudo-tensor $\overline{\tau}^{\alpha\beta}$ (the overbar means the PN expansion),
\begin{equation}\label{eq:ILsource}
	\dI_L = \FP \int \dd^d\mathbf{x} \,\Bigl(\frac{r}{r_0}\Bigr)^B \hat{x}_L \bigl[\overline{\tau}^{00} + \overline{\tau}^{ii}\bigr]+ \cdots\,,
\end{equation}
where $\hat{x}_L$ is the STF projection of $x_L = x_{i_1} ... x_{i_{\ell}}$, the finite part (FP) denotes a particular IR regularization when $B\to 0$ depending on an arbitrary regularization scale $r_0$, and the ellipsis denote other terms, known to all orders for general systems.

Once the external MPM metric is determined, we expand it to leading order in the distance between source and observer, to obtain the asymptotic waveform and the GW observables at infinity. Because harmonic coordinates asymptotically exhibit logarithms of the radial coordinate, which spoil the multipolar structure of the asymptotic waveform, we introduce a (non harmonic) ``radiative'' coordinate system adapted to the fall-off of the metric at infinity. By radiative coordinate system we mean one such that the retarded time $u=t-r$ is a null (or asymptotically null) coordinate, thus avoiding the logarithmic deviation of the retarded time in harmonic coordinates with respect to the true light cone. Alternatively, a different MPM construction defined in~\cite{B87}, corrects the coordinate light-cones at every PM order so as to iteratively build the null coordinate $u$, with the far zone logarithms therefore automatically cancelled. This construction has played a crucial role in our calculation of the 4PN flux.

The transverse-traceless (TT) waveform can be uniquely decomposed into two sets of STF ``radiative'' multipole moments $\dU_L$, $\dV_L$, which encode all the information about the asymptotic metric. These moments can straightforwardly be expressed in terms of the spherical harmonic modes as in Eq.~\eqref{eq:modeDecomposition}, see~(2.5)--(2.7) of~\cite{BFIS08}. The MPM construction determines (in principle up to any order) their relation to the canonical moments. For instance the leading term is the well-known nonlocal quadrupole tail integral arising at the 1.5PN order,
\begin{align}\label{eq:canradiative}
	& \dU_{ij}(u) = \dM^{(2)}_{ij}(u) + 2 \dM \int_{-\infty}^u \dd \tau \,\dM^{(4)}_{ij}(\tau) \nn\\*&\qquad\qquad\qquad\times \left[ \ln\left(\frac{u-\tau}{2 b_0}\right) + \frac{11}{12} \right] + \calO(5)\,,
\end{align}
where $\dM$ is the monopole of the multipole expansion (i.e. the Arnowitt-Deser-Misner mass of the spacetime) and $b_0$ is an arbitrary constant length scale linked to the definition of radiative coordinates. Beyond the tail term in~\eqref{eq:canradiative}, one finds the nonlinear memory effect at orders 2.5PN and 3.5PN; the tail-of-tail at 3PN order (a cubic interaction between two masses $\dM$ and the time-varying quadrupole); the tails-of-memory at 4PN order, which is a cubic interaction between $\dM$ and two varying quadrupoles~\cite{TB23}; the spin quadrupole tail at 4PN order (interaction $\dM\times\dS_i\times \dM_L$); and at 4.5PN order the quartic tail-of-tail-of-tail interaction, composed of three masses $\dM$ and the quadrupole~\cite{MBF16}.

The PN-MPM formalism is applied to compact binary sources. The compact objects are modelled by point particles (we neglect spins and other finite size effects). Furthermore, in a first stage of the calculation the two masses $m_i$ are considered to be constant, i.e. we neglect the black hole absorption, which should be added separately at the end of the PN calculation. 

The first manifestation of BH absorption is that the individual masses $m_i$ evolve according to \eqref{eq:mdot}.
We know that the horizon flux ${\cal F}_{\mathcal{H}_i}$ for nonspinning BHs is comparable to a 4PN orbital effect (beyond the 2.5PN Einstein quadrupole formula). For circular orbits it reads
\begin{equation}\label{eq:horizonflux}
	{\cal F}_{\mathcal{H}_i} = \frac{32}{5}\,\Bigl(\frac{m_i}{\mtot}\Bigr)^2 x^5 \Bigl[ \kappa \,x^4 + \calO(x^{5})\Bigr]\,,
\end{equation}
where $x$ is the orbital frequency~\eqref{eq:defx} and $\kappa=1$ has been computed in the test mass limit in Ref.~\cite{PS95}. By integrating Eqs.~\eqref{eq:mdot}--\eqref{eq:horizonflux} using the dominant radiation reaction effect on the orbit ($\dot{x}\propto x^5$) one finds
\begin{equation}\label{eq:solutionmi}
	m_i = \bar{m}_i + \frac{\kappa}{10} \mtot \nu \,x^5 + \calO(x^6)\,,
\end{equation}
where $\bar{m}_i$ denotes the initial mass before the GW emission (say, when $x\to 0$). Thus the evolution of the mass by BH absorption is comparable to a very small 5PN orbital effect. We can ignore this effect and consider the masses to be constant in our 4.5PN calculation. The second impact of BH absorption is that the energy flux balance law is modified as
\begin{equation}\label{eq:balanceBHabsorption}
	\frac{\dd E}{\dd t} = - \mathcal{F} - \sum_{i=1,2}\mathcal{F}_{\mathcal{H}_i}\,,
\end{equation} 
where ${\cal F}$ denotes the energy flux at infinity and $E$ is the binding energy. Therefore, when solving Eq.~\eqref{eq:balanceBHabsorption} to obtain the expression of the frequency evolution (or chirp), i.e. $\dot{x}$ expressed in terms of $x$, we shall obtain extra 4PN corrections due to the horizon fluxes. Those will affect the equations of motion at 4PN order beyond the dominant 2.5PN radiation reaction, i.e. corresponding to 6.5PN radiation reaction terms. When obtaining the 4.5PN flux at infinity, Eq.~\eqref{eq:Flux} below, we only required the expression of the frequency chirp at 1.5PN order beyond the dominant order (see (5.3b) of \cite{BFHLT23b}), hence the above 6.5PN radiation reaction terms can be safely discarded in our calculation. Our conclusion is that the BH absorption plays no role in the derivation of the energy flux at infinity~\eqref{eq:Flux} at this order, so we shall directly compare it with the GSF result in Sec.~\ref{sec:comparison}.

The equations of motion of the compact binary system have already been obtained at 4PN order~\cite{DJS14,DJS16,BBBFMc,MBBF17,FS19,FPRS19} and we extensively use the result. The most important task is the computation of the source multipole moments $\dI_L$, $\dJ_L$ for the binary source, see Eq.~\eqref{eq:ILsource}. The source moments represent the seed of the full construction through the steps~\eqref{eq:cansource}--\eqref{eq:canradiative}. The 4PN mass quadrupole moment $\dI_{ij}$ has been obtained in~\cite{MHLMFB20,LHBF22,LBHF22}. Previous works had determined that UV divergences, due to the modelling of compact objects by point masses, appear at the 3PN order and require the use of dimensional regularization. After regularization a shift of the particle's world lines permits to absorb the divergences. The works~\cite{MHLMFB20,LHBF22,LBHF22} showed that at the 4PN order IR divergences appear as well. For reasons explained in~\cite{BBBFMc,MBBF17}, we have used a variant of dimensional regularization called the ``$B\epsilon$'' regularization, where the finite part at $B=0$ in Eq.~\eqref{eq:ILsource} is applied on the top of the calculation performed in $d=3+\epsilon$ dimensions. An important point is that the regularization constant $r_0$ in~\eqref{eq:ILsource} is finally cancelled in the radiative moment~\eqref{eq:canradiative}, due notably to the $r_0$ dependence of the tails-of-memory. Besides the 4PN mass quadrupole moment, we also need the 3PN mass octupole $\dI_{ijk}$ and 3PN current quadrupole $\dJ_{ij}$, which have been obtained in~\cite{FBI15} and~\cite{HFB21}, respectively.

It is known that the GW half-phase $\psi$, defined by the decomposition of Eq. \eqref{eq:hlmAmplitudePhase}, and the orbital phase $\phi$, which is used in the computation to parametrize the circular motion of the binary, differ by a logarithmic phase modulation at order 1.5PN, which is due to the propagation of tails in the wave zone~\cite{Wi93,BS93,BIWW96},
\begin{equation}\label{eq:tailmodulation}
	\psi \equiv \phi - 2 \dM \ln\left(\frac{\omega}{\omega_0}\right) + \calO(5)\,.
\end{equation}
The constant $\omega_0$ is related to the constant $b_0$ in~\eqref{eq:canradiative} by $\omega_0^{-1} = 4b_0 \,\de^{\gamma_\text{E}-11/12}$ (where $\gamma_\text{E}$ is the Euler constant). While the GW phase $\psi$ and the corresponding frequency $\omega=\dd\psi/\dd u$ are directly measurable, the orbital phase $\phi$ can only be inferred via the theoretical prediction~\eqref{eq:tailmodulation}.

Taking the time derivative of Eq.~\eqref{eq:tailmodulation}, and using the fact that the time evolution of the phase occurs on a 2.5PN radiation reaction time scale, we find that the orbital frequency $\omega_\text{orb} = \dd \phi / \dd u$ and GW frequency $\omega = \dd \psi / \dd u$ differ by a small 4PN term, which thus enters for the first time in the 4PN waveform. A consequence is that the frequency parameter~$x$ defined by~\eqref{eq:defx} differs from its counterpart $y= (m_\text{tot} \,\omega_\text{orb})^{2/3}$ by a small 4PN correction:
\begin{equation}\label{xy}
	x = y\,\biggl\{ 1 - \frac{192}{5} \,\nu \, y^4 \biggl[\ln\biggl(\frac{y}{y_0}\biggr) + \frac{2}{3} \biggr]  + \calO(y^5) \biggr\}\,,
\end{equation}
where $y_0 = (m_\text{tot}\,\omega_0)^{2/3}$ and $\calO(y^5)=\calO(10)$. When expressing physical results in terms of the GW observables $\psi$ and $\omega$, the arbitrary scale $b_0$ is cancelled out (see~\cite{BFHLT23b,T25_Schott} for details). 

Another subtlety that arises at 4PN order is the treatment of the tail integral~\eqref{eq:canradiative}. To compute it explicitly at 3.5PN order for a quasicircular orbit, it is sufficient to assume that the orbital frequency and orbit radius are constant --- this is the ``adiabatic'' approximation. However this approximation is no longer valid at relative 2.5PN precision, because this is the order of radiation reaction; one must then consistently account for the time evolution of the orbital frequency as well. Since the tail term enters at the 1.5PN order, the first ``post-adiabatic'' correction will affect the waveform at the 4PN order, and it needs to be properly evaluated in order to control the 4PN flux and modes~\cite{BFHLT23b}.

The end result for the dominant mode~\mbox{$(\ell,m)=(2,2)$} at 4PN order, following the definition~\eqref{eq:modeDecomposition}--\eqref{eq:hlmAmplitudePhase}, reads
\begin{align}\label{eq:h22}
	& \hat{h}_{22} = \sqrt{\frac{64\pi}{5}}\,\nu x\bigg\{ 1 + \Bigl(-\tfrac{107}{42}+\tfrac{55}{42}\nu\Bigr) x 
	+2 \pi x^{3/2} 
	\nn\\
	&+ \Bigl(-\tfrac{2173}{1512}-\tfrac{1069}{216}\nu+\tfrac{2047}{1512}\nu^2\Bigr)x^2
	+  \Bigl(-\tfrac{107}{21} + \tfrac{34}{21} \nu\Bigr) \pi x^{5/2} \nn \\
	&+ \biggl[\tfrac{27027409}{646800}-\tfrac{856}{105}\,\gamma_\text{E} +\tfrac{2 \pi ^2}{3} + \Bigl(-\tfrac{278185}{33264}+\tfrac{41 \pi^2}{96}\Bigr) \nu \nn\\*
	&\qquad -\tfrac{20261}{2772}\nu^2+\tfrac{114635}{99792}\nu^3-\tfrac{428}{105} \ln (16 x)\biggr] x^3
	\nn \\
	&+  \Bigl( -\tfrac{2173}{756} -\tfrac{2495}{378} \nu + 
	\tfrac{40}{27} \nu^2 \Bigr) \pi x^{7/2} 
	\nn \\
	&+ \biggl[- \tfrac{846557506853}{12713500800} + \tfrac{45796}{2205}\gamma_\text{E} - \tfrac{107}{63}\pi^2 + \tfrac{22898}{2205}\ln(16x) \nn \\
	&  +\Bigl(- \tfrac{336005827477}{4237833600} + \tfrac{15284}{441}\gamma_\text{E}  - \tfrac{9755}{32256}\pi^2 + \tfrac{7642}{441}\ln(16x)\Bigr)\nu \nn \\
	&+\Bigl( \tfrac{256450291}{7413120} - \tfrac{1025}{1008}\pi^2 \Bigr)\nu^2 - \tfrac{81579187}{15567552}\nu^3 + \tfrac{26251249}{31135104}\nu^4 \biggr] x^4 \nn \\
	&+	\calO(x^{9/2}) \bigg\} \,.
\end{align}
Unlike in previous works, e.g.~\cite{BFHLT23a,BFHLT23b}, we have chosen the convention that the amplitude of the (2,2) mode be real valued. The amplitude given in Eq.~\eqref{eq:h22} is thus equal (up to a global prefactor) to the modulus of the amplitude given by Eq.~(11) of~\cite{BFHLT23a}. This difference of course leads to corrections to the phase evolution~$\psi(u)$, but these are very small 5PN modulations beyond the leading order in the phase (which is $\propto x^{-5/2}$), for instance comparable to neglected terms in Eq.~\eqref{xy}. The result~\eqref{eq:h22} is in perfect agreement with linear black-hole perturbation theory at first order (i.e. 1SF) in the mass ratio~\cite{TSasa94,TTS96}. For the other modes up to 3.5PN order, we refer to Refs.~\cite{BFIS08,H23}.

Finally, adding up all pieces together we obtain the quasi-circular 4.5PN flux~\cite{BFHLT23a,BFHLT23b}
\begin{align}\label{eq:Flux}
	& \mathcal{F} = \frac{32}{5}\nu^2 x^5\biggl\{
	1 
	+ \Bigl(-\tfrac{1247}{336} - \tfrac{35}{12}\nu \Bigr) x 
	+ 4\pi x^{3/2}
	\nn\\
	& 
	+ \Bigl(-\tfrac{44711}{9072} +\tfrac{9271}{504}\nu + \tfrac{65}{18} \nu^2\Bigr) x^2 
	+ \Bigl(-\tfrac{8191}{672}-\tfrac{583}{24}\nu\Bigr)\pi x^{5/2}
	\nn\\
	& 
	+ \biggl[\tfrac{6643739519}{69854400}+ \tfrac{16}{3}\pi^2-\tfrac{1712}{105}\gamma_\text{E} - \tfrac{856}{105} \ln (16\,x) 
	\nn\\
	&~~ + \Bigl(-\tfrac{134543}{7776} + \tfrac{41}{48}\pi^2 \Bigr)\nu 
	- \tfrac{94403}{3024}\nu^2 
	- \tfrac{775}{324}\nu^3 \biggr] x^3 
	\nn\\
	&
	+ \Bigl(-\tfrac{16285}{504} + \tfrac{214745}{1728}\nu +\tfrac{193385}{3024}\nu^2\Bigr)\pi x^{7/2} 
	\nn\\
	&
	+ \biggl[ -\tfrac{323105549467}{3178375200} + \tfrac{232597}{4410}\gamma_\text{E} - \tfrac{1369}{126} \pi^2 
	\nn\\
	&~~ + \tfrac{39931}{294}\ln 2 - \tfrac{47385}{1568}\ln 3 + \tfrac{232597}{8820}\ln x   
	\nn\\
	&~~
	+ \Bigl( -\tfrac{1452202403629}{1466942400} + \tfrac{41478}{245}\gamma_\text{E} - \tfrac{267127}{4608}\pi^2  
	\nn\\
	&\quad\quad + \tfrac{479062}{2205}\ln 2 + \tfrac{47385}{392}\ln 3  + \tfrac{20739}{245}\ln x \Bigr)\nu
	\nn\\
	&~~
	+ \Bigl( \tfrac{1607125}{6804} - \tfrac{3157}{384}\pi^2 \Bigr)\nu^2 + \tfrac{6875}{504}\nu^3 + \tfrac{5}{6}\nu^4 \biggr] x^4
	\nn\\ 
	& 
	+ \biggl[ \tfrac{265978667519}{745113600} - \tfrac{6848}{105}\gamma_\text{E} - \tfrac{3424}{105} \ln (16 \,x)
	\nn\\
	&~~ + \Bigl( \tfrac{2062241}{22176} + \tfrac{41}{12}\pi^2 \Bigr)\nu 
	- \tfrac{133112905}{290304}\nu^2 - \tfrac{3719141}{38016}\nu^3 \biggr] \pi x^{9/2} 
	\nn\\
	&
	+ \calO(x^5) \biggr\}\,.
\end{align}
The decomposition of the total flux into mode contributions $\mathcal{F}_{\ell \textrm{m}}$ is relegated to Appendix~\ref{app:Flm}.

\section{Review of 2SF fluxes}
\label{sec:2SF}

In the GSF formulation of the binary problem, the primary object is taken to be a black hole, the secondary object is reduced to a point particle, and the spacetime metric is expanded in powers of the binary's mass ratio.

We denote the zeroth-order, background metric as $\mathring{g}_{\alpha\beta}(x^i)$, representing the spacetime of the primary black hole in isolation, meaning a Schwarzschild geometry with constant mass $\mathring{m}_1$. The variables $x^i=(r/\mathring{m}_1,\theta,\phi)$ are the usual Schwarzschild spatial coordinates, adimensionalized with the background mass $\mathring{m}_1$. We define the mass ratio $\varepsilon$ in terms of this background mass as $\varepsilon\equiv m_2/\mathring{m}_1$.

The first- and second-order corrections to $\mathring{g}_{\alpha\beta}$, due to the orbiting particle, are obtained using a multiscale formulation of the expansion in $\varepsilon$, detailed in Refs.~\cite{MP21,Miller:2023ers}. In this approach, powers of $\varepsilon$ represent \emph{post-adiabatic} (PA) orders~\cite{Hinderer:2008dm,
Pound2022black}; the leading (0PA) order is the traditional adiabatic approximation, in which the system adiabatically evolves through a smooth sequence of test-particle orbits. Concretely, our multiscale approach assumes the spacetime's evolution arises entirely from the evolution of the binary's mechanical variables $(\phi_p,\Omega,\delta m_1)$, where $\phi_p$ is the particle's orbital (azimuthal) phase, 
\begin{equation}
\Omega \equiv \mathring{m}_1 \frac{\dd\phi_p}{\dd t}
\end{equation}
is its (adimensionalized) slowly evolving orbital frequency, and
\begin{equation}
\delta m_1\equiv (m_1-\mathring{m}_1)/m_2
\end{equation}
is an adimensionalized correction to the black hole's mass parameter, which evolves due to the black hole's absorption of radiation.\footnote{The Einstein equations dictate that a nonzero spin $\delta s_1$ also arises due to absorption of angular momentum, and our complete 2SF flux calculations include this correction. However, to compare with PN results, we artificially set it to zero.} The mass evolves by an amount of order $m_2$ over the course of the inspiral, and we adimensionalize $\delta m_1$ using $m_2$ to make it order unity. The mass $m_2$, on the other hand, is constant at 2SF order. 

In terms of these mechanical variables and the spatial coordinates $x^i$, the metric is expanded as
\begin{multline}
\label{eq:metric}
g_{\alpha\beta} = \mathring{g}_{\alpha\beta}(x^i) + \varepsilon h^{(1)}_{\alpha\beta}(\phi_p,\Omega,\delta m_1, x^i)\\ 
+ \varepsilon^2 h^{(2)}_{\alpha\beta} (\phi_p,\Omega,\delta m_1, x^i) +{\cal O}(\varepsilon^3)\,.
\end{multline}
Because the mass only changes by an amount $\sim m_2$ over the inspiral time $\sim 1/\varepsilon$, the evolving correction is treated perturbatively rather than altering $\mathring{g}_{\alpha\beta}$. No explicit dependence on a time coordinate appears in the metric~\eqref{eq:metric} because time dependence is subsumed into the dependence on $(\phi_p,\Omega,\delta m_1)$. 

As described in Refs.~\cite{MP21,Miller:2023ers}, the mechanical variables are treated as functions of a hyperboloidal time coordinate $s$ that reduces to Schwarzschild time $t$ on the particle, retarded Eddington-Finkelstein time $u$ at future null infinity, and advanced Eddington-Finkelstein time $v$ at the horizon, adimensionalized by $\mathring{m}_1$. Their time evolution (and hence, the spacetime's evolution) is governed by equations of the form 
\begin{subequations}\begin{align}
\frac{\dd\phi_p}{\dd s} &= \Omega\,,\label{phidot}\\
\frac{\dd\Omega}{\dd s} &= \varepsilon\Bigl[F_0^\Omega(\Omega) + \varepsilon F_1^\Omega(\Omega,\delta m_1)+{\cal O}(\varepsilon^2)\Bigr]\,,\label{eq:Omegadot}\\
\frac{\dd \delta m_1}{\dd s} &= \varepsilon\mathcal{F}^{(1)}_{{\cal H}_1}(\Omega) +{\cal O}(\varepsilon^2)\,.\label{eq:m1dot}
\end{align}\end{subequations}
Here $\mathcal{F}^{(1)}_{{\cal H}_1}$ is the standard energy flux carried by $h^{(1)}_{\alpha\beta}$ across the primary's event horizon~\cite{Hughes:1999bq}. The forcing function $F_0^\Omega(\Omega)$ is the standard adiabatic (0PA) rate of change of the orbital frequency, which is related to the emitted energy flux by 
\begin{equation}
F_0^\Omega = -\frac{\mathcal{F}^{(1)}+\mathcal{F}^{(1)}_{{\cal H}_1}}{\dd{\cal E}_0/\dd\Omega}\,,\label{F0Omega}
\end{equation}
where $\mathcal{F}^{(1)}$ is the standard energy flux carried by $h^{(1)}_{\alpha\beta}$ to future null infinity~\cite{Hughes:1999bq}, and ${\cal E}_0$ is the specific orbital energy of a test mass on a circular geodesic with frequency $\Omega$. This forcing function $F_0^\Omega$ enters the Einstein field equations for $h^{(2)}_{\alpha\beta}$ and into explicit expressions for the 2SF flux below; the first post-adiabatic (1PA) forcing function $F_1^\Omega$, on the other hand, does not enter into the calculation of the 2SF fluxes. Note that here we have factored out powers of $\varepsilon$ such that $F_n^\Omega$, ${\cal F}^{(1)}_{{\cal H}_1}$, ${\cal F}^{(1)}$, and ${\cal E}_0$ are all $\varepsilon$-independent functions of $\Omega$.

The waveform in the multiscale expansion, in analogy with Eq.~\eqref{eq:metric}, takes the form
\begin{equation}\label{eq:2SF waveform}
h_{\ell m} = \mathring{m}_1\left[\varepsilon h^{(1)}_{\ell m}(\Omega)+\varepsilon^2 h^{(2)}_{\ell m}(\Omega,\delta m_1)+{\cal O}(\varepsilon^3)\right]\de^{-\di m\phi_p}\,,
\end{equation}
where $h^{(n)}_{\ell m}$ is obtained from the coefficient of $1/r$ in the large-$r$ limit of the metric perturbation $h^{(n)}_{\alpha\beta}$  in a Bondi-Sachs (radiative) gauge~\cite{Madler:2016xju}. In practice, we solve the Einstein equations for $h^{(n)}_{\alpha\beta}$ in the Lorenz gauge,\footnote{Here the Lorenz gauge condition is $\mathring{g}^{\beta\gamma}\mathring{\nabla}_\beta\bar h_{\alpha\gamma}=0$, where $\mathring{g}^{\beta\gamma}$ is the inverse of the background metric $\mathring{g}_{\beta\gamma}$, $\mathring{\nabla}_\beta$ is the covariant derivative compatible with $\mathring{g}_{\alpha\beta}$, and \mbox{$\bar h_{\alpha\beta}\equiv(\mathring{g}_{\alpha}{}^{\mu}\mathring{g}_{\beta}{}^\nu-\frac{1}{2}\mathring{g}_{\alpha\beta}\mathring{g}^{\mu\nu}) (g_{\mu\nu}-\mathring{g}_{\mu\nu})$}. This gauge shares many essential features with the harmonic coordinate condition $g^{\alpha\beta}\Gamma^\gamma_{\alpha\beta}=0$ used in PN theory.} which is not well behaved at large $r$~\cite{Pound:2015wva,MP21,Miller:2023ers} (as in the PN case in Sec.~\ref{sec:PN}), and then transform to a Bondi-Sachs gauge to eliminate the ill-behaved pieces of the Lorenz-gauge solution. 

The transformation to the Bondi-Sachs gauge is explained in Ref.~\cite{Cunningham:2024dog}. In that reference, we also discovered an unforeseen contribution to $h^{(2)}_{\ell m}$, due to the interaction between oscillatory modes and gravitational-memory modes in $h^{(1)}_{\alpha\beta}$. The new ``memory distortion'' term contributes to the flux ${\cal F}^{(2)}$, but we do not include it in our comparisons in this paper because (i) numerical results for it have only recently been obtained and are not yet published, and (ii) the contribution is numerically very small and appears to enter only at 5PN~\cite{Cunningham:Private}.

It is possible to obtain the flux directly from the waveform~\eqref{eq:2SF waveform}. Explicitly, substituting Eq.~\eqref{eq:2SF waveform} into the flux formula~\eqref{eq:defFlm}, one finds
\begin{align}
	\mathcal{F}_{\ell m} &= \frac{1}{8\pi}\biggl\{\varepsilon^2 \bigl|m \Omega \,h^{(1)}_{\ell m}\bigr|^2+2\varepsilon^3 \,\Re \left(m^2\Omega^2 h^{(2)}_{\ell m}h^{(1)*}_{\ell m}\right.\nonumber\\
	&\qquad\qquad\left. + \di m\Omega \,F^\Omega_0 \,h^{(1)*}_{\ell m}\partial_\Omega h^{(1)}_{\ell m}\right) + {\cal O}(\varepsilon^4) \biggr\}\,.\label{eq:GSF_flux}
\end{align}
To recast this in a form suitable for comparison with the PN flux, we can re-express it in terms of the common variables $(x,\mtot,\nu)$ and re-expand in powers of $\nu$ at fixed $(x,\mtot)$. 

However, before proceeding to the flux, it will be instructive to instead re-express the waveform itself in terms of the shared variables. This will provide a more intuitive link between the waveform and the flux. We start by expressing the waveform in terms of $(\mtot,\nu)$ and an orbital frequency parameter (as opposed to the waveform frequency). The various masses are related by  
\begin{subequations}\begin{align}
\mathring{m}_1 &= m_1-m_2\delta m_1\,,\\
m_1 &=\frac{\mtot}{2} \bigl(1+\sqrt{1-4\nu}\bigr)\,,\\
m_2 &= \frac{\mtot}{2}\bigl(1-\sqrt{1-4 \nu}\bigr)\,.
\end{align}\end{subequations}
Re-expanding~\eqref{eq:2SF waveform} in powers of $\nu$ at fixed $\mtot$ leads to 
\begin{equation}\label{1PAT1 hlm - nu}
h_{\ell m} = \mtot\left[\nu \tilde h^{(1)}_{\ell m}(y)+\nu^2\tilde h^{(2)}_{\ell m}(y)+{\cal O}(\nu^3)\right]\de^{-\di m\phi_p}\,,
\end{equation}
where 
\begin{subequations}\label{htilde}\begin{align}
\tilde h^{(1)}_{\ell m}(y) &= h^{(1)}_{\ell m}(y^{3/2})\,,\\
\tilde h^{(2)}_{\ell m}(y) &= h^{(2)}_{\ell m}(y^{3/2},\delta m_1) + h^{(1)}_{\ell m}(y^{3/2})\nonumber\\
&\quad -(1+\delta m_1)y^{3/2} \partial_{\Omega}h^{(1)}_{\ell m}(y^{3/2})\,. \label{h2tilde}
\end{align}\end{subequations}
Here $y$ is the parameter introduced above Eq.~\eqref{xy}; it is related to our dimensionless $\Omega$ by 
\begin{equation}
y=\left(\frac{\mtot}{\mathring{m}_1}\Omega\right)^{2/3}\,. 
\end{equation}
Note that the dependence on $\delta m_1$ cancels on the right-hand side of Eq.~\eqref{h2tilde}, leaving only a dependence on the frequency variable $y$.\footnote{We emphasize that this only represents the elimination of the split of the physical, evolving mass $m_1$ into a constant background mass and an evolving correction. The mass's evolution enters directly in the waveform's time dependence via Eq.~\eqref{eq:m1dot}. The rate of change of the mass also enters the $\ell=0$ field equations for $h^{(2)}_{\alpha\beta}$ when time derivatives  act on the metric perturbation~\eqref{eq:metric}~\cite{MP21}, but the resulting contribution to the $\ell=0$ mode only enters the $\mathcal{O}(\nu^4)$ flux. Finally, the horizon absorption enters into the field equations and the $\mathcal{O}(\nu^3)$ flux to infinity via $F^\Omega_0$ in Eq.~\eqref{F0Omega}, but $F^\Omega_0$ can equivalently be calculated from the local first-order dissipative self-force rather than from fluxes.}

To express the waveform in terms of the waveform frequency, we go one step further by writing the waveform~\eqref{1PAT1 hlm - nu} in terms of a real amplitude and complex phase factor, 
\begin{equation}\label{eq:real amplitudes}
h_{\ell m} = \mtot A_{\ell m}\de^{-\di m\psi_{\ell m}}\,,
\end{equation}
where $A_{\ell m}=|h_{\ell m}|/\mtot$ and $m\psi_{\ell m}=-\arg(h_{\ell m})$. This corresponds to Eq.~\eqref{eq:hlmAmplitudePhase} with the identifications $\psi=\psi_{22}$
and $A_{\ell m}=|\hat h_{\ell m}|$. 
Substituting Eq.~\eqref{1PAT1 hlm - nu}, we then find the analogue of the PN phase modulation~\eqref{eq:tailmodulation}:
\begin{align}\label{eq:psilm}
    m\psi_{\ell m} &= m\phi_p -\nu^0\arg\bigl(\tilde h^{(1)}_{\ell m}\bigr)\nonumber\\*
 & +\nu\, \frac{\Re\,\tilde h^{(2)}_{\ell m}\,\Im\,\tilde h^{(1)}_{\ell m}- \Im\,\tilde h^{(2)}_{\ell m}\,\Re\,\tilde h^{(1)}_{\ell m}}{\bigl|\tilde h^{(1)}_{\ell m}\bigr|^2} +{\cal O}(\nu^2)\,.
\end{align}
We can assess the importance of each term on the right-hand side by noting that the phases $\psi_{\ell m}$ and $\phi_p$ are both ${\cal O}(\nu^{-1})$ quantities; on the radiation-reaction timescale $\mtot/\nu$, the phases evolve by an amount of order $1/\nu$. Hence the order-$\nu^0$ term in Eq.~\eqref{eq:psilm} represents a relative 1PA phase correction, while the order-$\nu$ term represents a relative 2PA phase correction. Likewise, the order-$\nu$ term will only affect the waveform frequency at 2PA order ($\sim~\nu^2$) because it only involves the slowly evolving quantities ($y,\mtot,\nu$), with no direct dependence on $\phi_p$. Hence, we only require the first two terms in Eq.~\eqref{eq:psilm}.

Finally, we relate the waveform frequency to the orbital frequency by applying a time derivative to Eq.~\eqref{eq:psilm} and using~\eqref{eq:Omegadot} together with $\dd\nu/\dd u=\mathcal{O}(\nu^3)$. This yields the result previously presented in Ref.~\cite{Albertini:2022rfe} (generalized to generic $\ell m$),
\begin{equation}\label{Omega to omega}
\dot\psi_{\ell m} \equiv \omega_{\ell m} = \dot\phi_p +\nu\omega^{(1)}_{\ell m} +\mathcal{O}(\nu^2)\,,
\end{equation}
with
\begin{equation}\label{omega1}
m\omega^{(1)}_{\ell m} = \frac{2F^\Omega_0}{3y^{1/2}\bigl|\tilde h^{(1)}_{\ell m}\bigr|^2}\left({\Im}\,\tilde h^{(1)}_{\ell m}\, {\Re}\,\partial_y\tilde h^{(1)}_{\ell m} - {\Re}\,\tilde h^{(1)}_{\ell m}\, {\Im}\,\partial_y\tilde h^{(1)}_{\ell m}\right)\!.
\end{equation}
Following the general discussion in the Introduction, we define the waveform frequency as $\omega=\omega_{22}$. The analogue of the PN equation~\eqref{xy} then reads
\begin{equation}
    x = y\left[1+\frac{2\mtot\omega^{(1)}}{3 y^{3/2}}\nu + {\cal O}(\nu^2)\right]\,,\label{eq:x vs y SF}
\end{equation}
with $\omega^{(1)}\equiv\omega^{(1)}_{22}$. We can also express the general $\omega_{\ell m}$ as a function of $\omega$:
\begin{equation}
    \omega_{\ell m} = \omega + \nu \Delta\omega^{(1)}_{\ell m} + {\cal O}(\nu^2)\,,
\end{equation}
where $\Delta\omega^{(1)}_{\ell m} \equiv \omega^{(1)}_{\ell m}-\omega^{(1)}$.

Substituting Eq.~\eqref{eq:x vs y SF} into Eq.~\eqref{1PAT1 hlm - nu}, we obtain $h_{\ell m}$ in terms of the waveform frequency:
\begin{equation}\label{eq:real amplitudes final}
h_{\ell m} = \mtot\Bigl[\nu A^{(1)}_{\ell m}(x)+\nu^2 A^{(2)}_{\ell m}(x)+{\cal O}(\nu^3)\Bigr]\de^{-\di m\psi_{\ell m}}\,,
\end{equation}
where
\begin{subequations}\begin{align}
    A^{(1)}_{\ell m} &= \bigl|\tilde h^{(1)}_{\ell m}\bigr|\,,\\
    A^{(2)}_{\ell m} &= \frac{1}{2\bigl|\tilde h^{(1)}_{\ell m}\bigr|}\biggl[\tilde h^{(1)*}_{\ell m}\biggl(\tilde h^{(2)}_{\ell m} -\frac{2\mtot\omega^{(1)}}{3x^{1/2}}\partial_x \tilde h^{(1)}_{\ell m}\biggr)+\text{c.c.}\biggr]\,. 
\end{align}\end{subequations}
Functions of $y$ or $\Omega$ on the right side are evaluated at $x$ or $x^{3/2}$, respectively.

In terms of the real amplitudes $A_{\ell m}$, the flux~\eqref{eq:defFlm} takes a simple form,
\begin{align}
    {\cal F}_{\ell m} &= \frac{\mtot^2}{8\pi}\biggl[m^2 \omega_{\ell m}^2 (A_{\ell m})^2 +  \left(\frac{\dd x}{\dd u}\right)^{\!\!2} (\partial_x A_{\ell m})^2+\mathcal{O}(\nu^4)\biggr]\nonumber\\
    &= \frac{m^2 \mtot^2 \omega_{\ell m}^2}{8\pi} (A_{\ell m})^2 +{\cal O}(\nu^4)\,,
\end{align}
where we have used $A_{\ell m}=\mathcal{O}(\nu)$ together with $d\nu/du=\mathcal{O}(\nu^3)$ and $\dd m_1/\dd u=\mathcal{O}(\nu^2)$ to neglect time derivatives of $\nu$ and $\mtot$ in the first line, and $\dd x/\dd u=\mathcal{O}(\nu)$ in the second line.
After substituting the expansions of $A_{\ell m}$ and $\omega_{\ell m}$, we can rewrite this in the form~\eqref{eq:GSF flux generic}:
\begin{multline}
     {\cal F}_{\ell m} = \frac{\nu^2m^2\mtot^2}{8\pi}\biggl\{\omega^2\bigl(A^{(1)}_{\ell m}\bigr)^2
     + 2\nu \Bigl[\omega^2 A^{(1)}_{\ell m}A^{(2)}_{\ell m}\\*
     +\omega\Delta\omega^{(1)}_{\ell m}\bigl(A^{(1)}_{\ell m}\bigr)^2\Bigr] 
     + {\cal O}(\nu^2)\biggr\}\,.
\end{multline}

\section{Comparison}
\label{sec:comparison}

\begin{figure}
    \centering
    \includegraphics[width=\columnwidth]{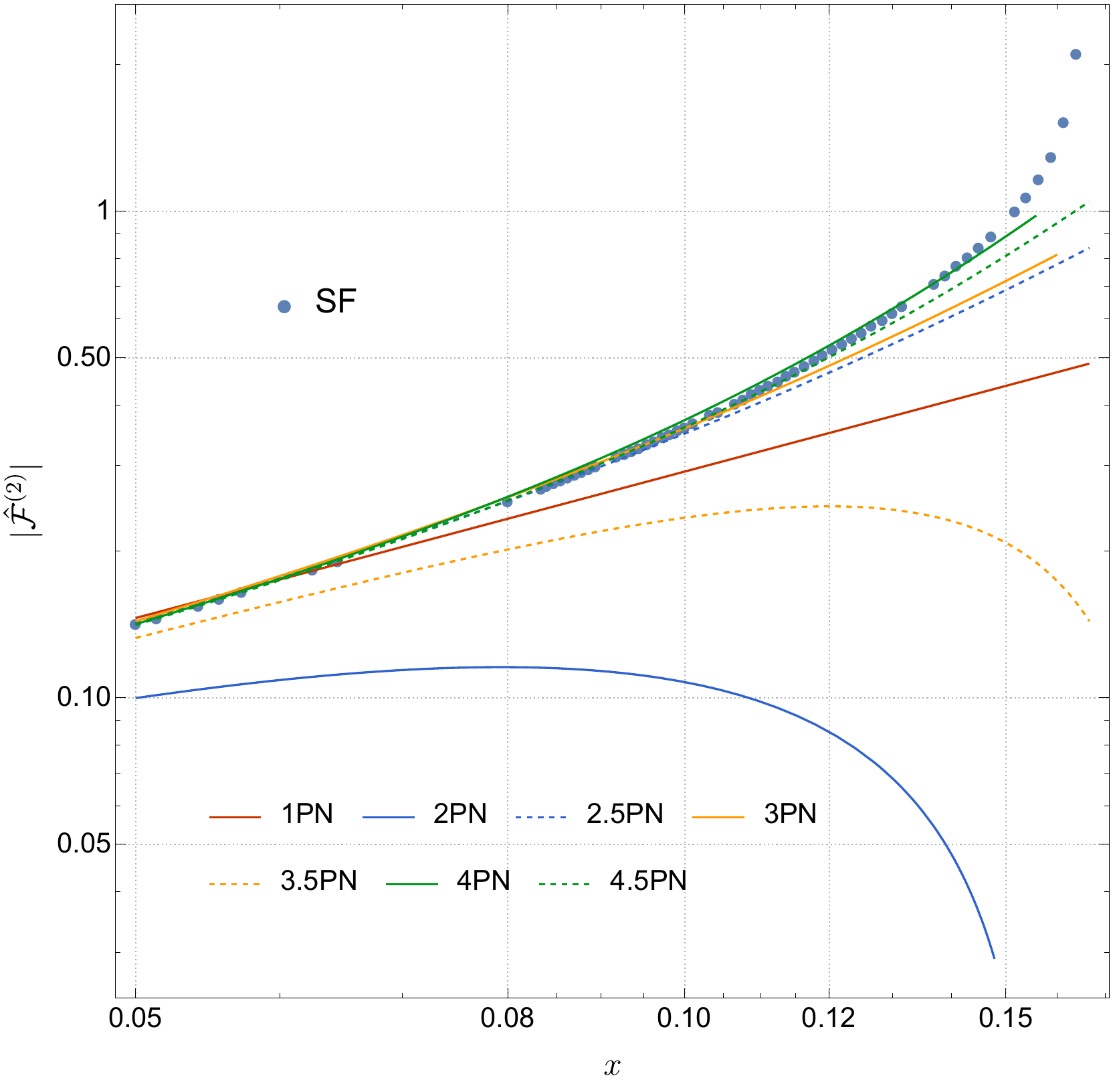}
    \caption{Comparison between GSF and PN for the 2SF $\mathcal{O}(\nu)$ contribution to the (Newtonian-normalized) total flux~\eqref{eq:normalizedflux}. The 2SF results are shown with (blue) dots. The PN results are shown with coloured curves; integer PN orders are shown with solid curves and corresponding half-integer orders are shown with dashed curves.}
    \label{fig:comparison_total_flux}
\end{figure}

\begin{figure}
    \centering
    \includegraphics[width=\columnwidth]{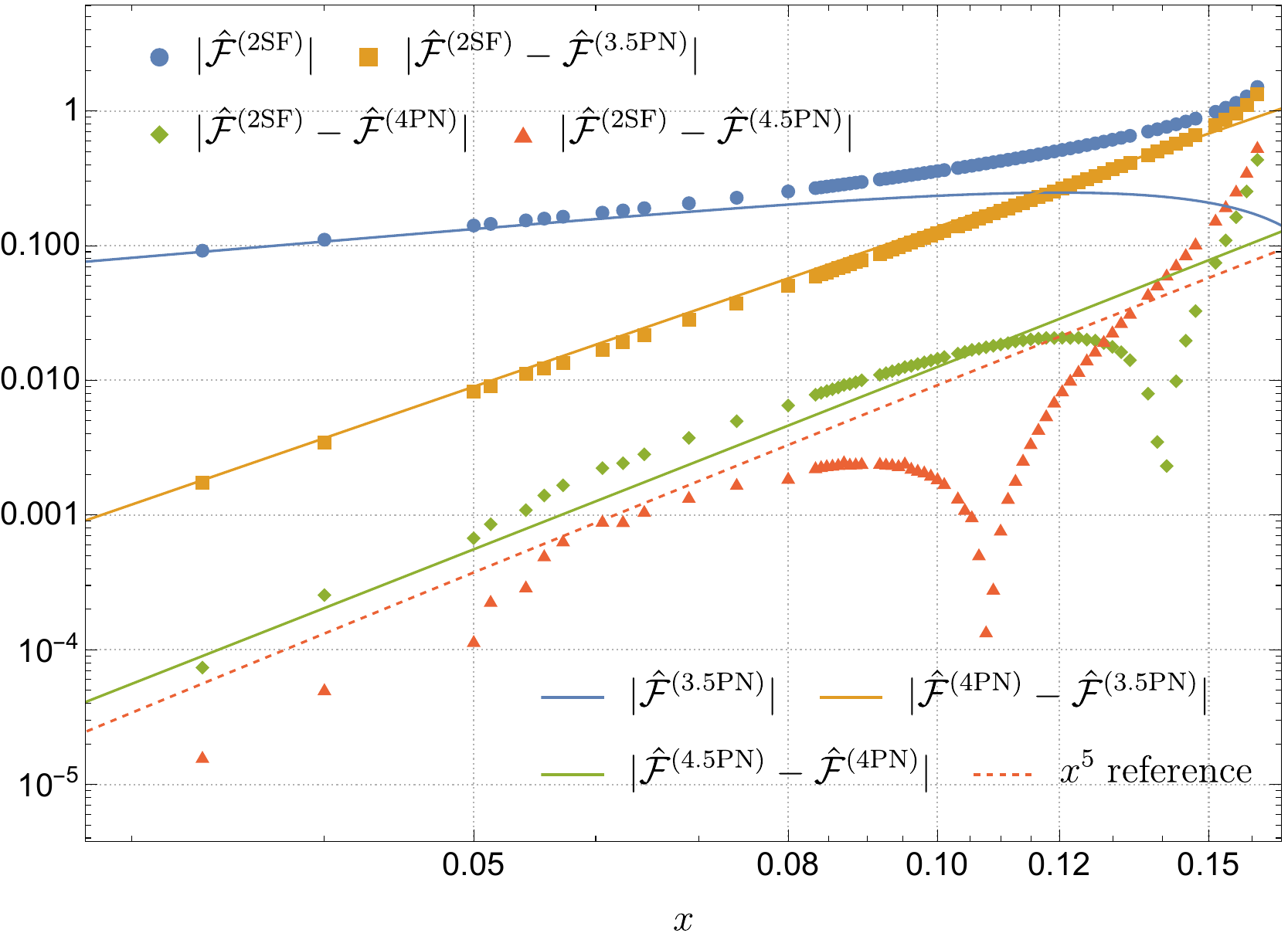}
    \caption{
    Detailed comparison of $\hat{\mathcal{F}}^{(2)}$, the $\mathcal{O}(\nu)$ coefficient of the (Newtonian-normalized) total flux, computed from (i) numerical GSF and (ii) PN theory. The 2SF data are shown by (blue) dots and the corresponding 3.5PN series is plotted as a solid (blue) curve. As ever higher-order PN series are subtracted from the 2SF results, the residuals are compared with the next term in the PN series at small $x$. After subtracting the 3.5PN series from the 2SF data, one gets the (yellow) squares which follow the 4PN term (yellow curve). Subtracting the 4PN series from the SF data, one gets the (green) diamonds, which are compared against the 4.5PN term (green curve). The amplitude of the residual with the 4PN series is consistent with the amplitude of the 4.5PN term, although the slope of the residual is unclear, partly due to numerical noise. Further subtracting the 4.5PN series gives the (red) triangles. For sufficiently small values of $x$, we would expect the residual to scale as $x^5$. We plot a (red) dashed reference $x^5$ curve here but the accuracy of our numerical data is insufficient to claim agreement. One reason that the agreement with 4.5PN is not clear can be seen by considering the dominant $(2,2)$ mode: over the frequency range of our numerical GSF data there is evidence that the residual with 4PN has not reached the asymptotic regime -- see Fig.~\ref{fig:flux22}. The agreement between PN and GSF results is clearer for some individual $(\ell,m)$ modes -- see Figs.~\ref{fig:flux22}, \ref{fig:comparisonmodes_l3m3_l3m2} and the figures in Appendix \ref{app:Flm-comparison}.
    }
    \label{fig:detailed_comparison_total_flux}
\end{figure}

\begin{figure}
    \centering
    \includegraphics[width=\columnwidth]{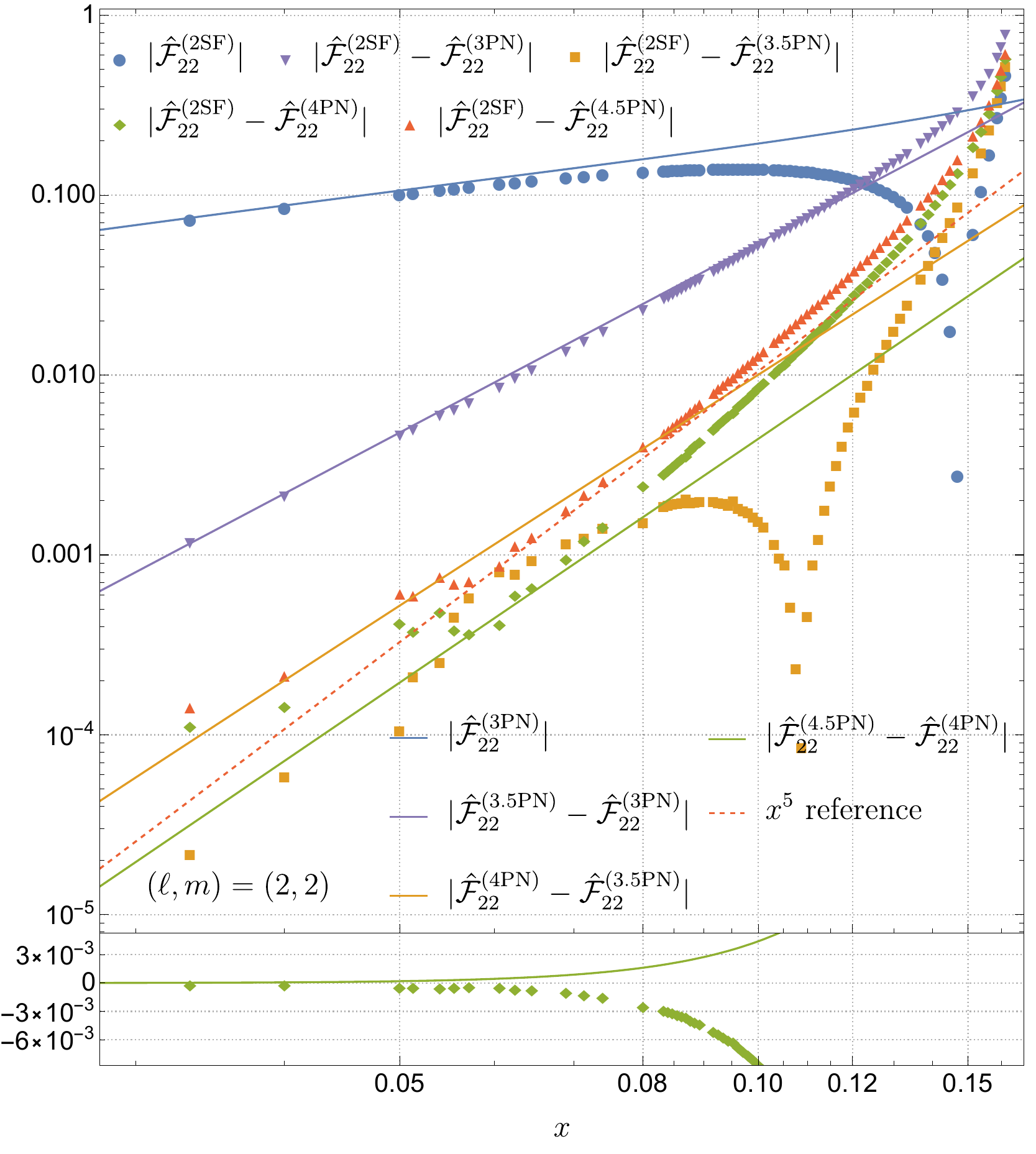}
    \caption{(\textit{Top}) Detailed comparison of $\hat{\mathcal{F}}^{(2)}_{22}$, the $\mathcal{O}(\nu)$ coefficient of the $(\ell,m)=(2,2)$ mode of the (Newtonian-normalized) flux, computed from (i) numerical GSF and (ii) PN theory.
    The conventions and colorings are the same as in Fig.~\ref{fig:detailed_comparison_total_flux}, but the blue curve now shows the 3PN series and the additional upside-down (purple) triangles show the residual after subtracting the 3PN series from the 2SF data. The latter residual closely follows the 3.5PN term (purple curve). The results of comparisons with higher PN orders are more ambiguous due to numerical noise in the GSF data. For example, the residual after subtracting the 4PN series from the 2SF result (green diamonds) appears at first sight to approach the 4.5PN term [before degrading again for even smaller values of $x$, due to numerical noise], but this is in fact an illusion caused by plotting the absolute magnitude of the residual. (\textit{Bottom}) A subset of the above data  (without taking the absolute magnitude) on a linear-log scale. Over the range of frequency values where we have numerical data, the residual between the GSF and 4PN is negative, but the 4.5PN term is positive. Thus, in order for these to agree asymptotically, there must be a zero crossing in the residual data at smaller values of $x$ than we have access to. Only after this zero crossing would we expect the residual and the 4.5PN curve to agree. The agreement between GSF and PN for other $(\ell,m)$ modes is clearer -- see Appendix \ref{app:Flm-comparison}.}
    \label{fig:flux22}
\end{figure}

\begin{figure*}
    \includegraphics[width=0.48\textwidth]{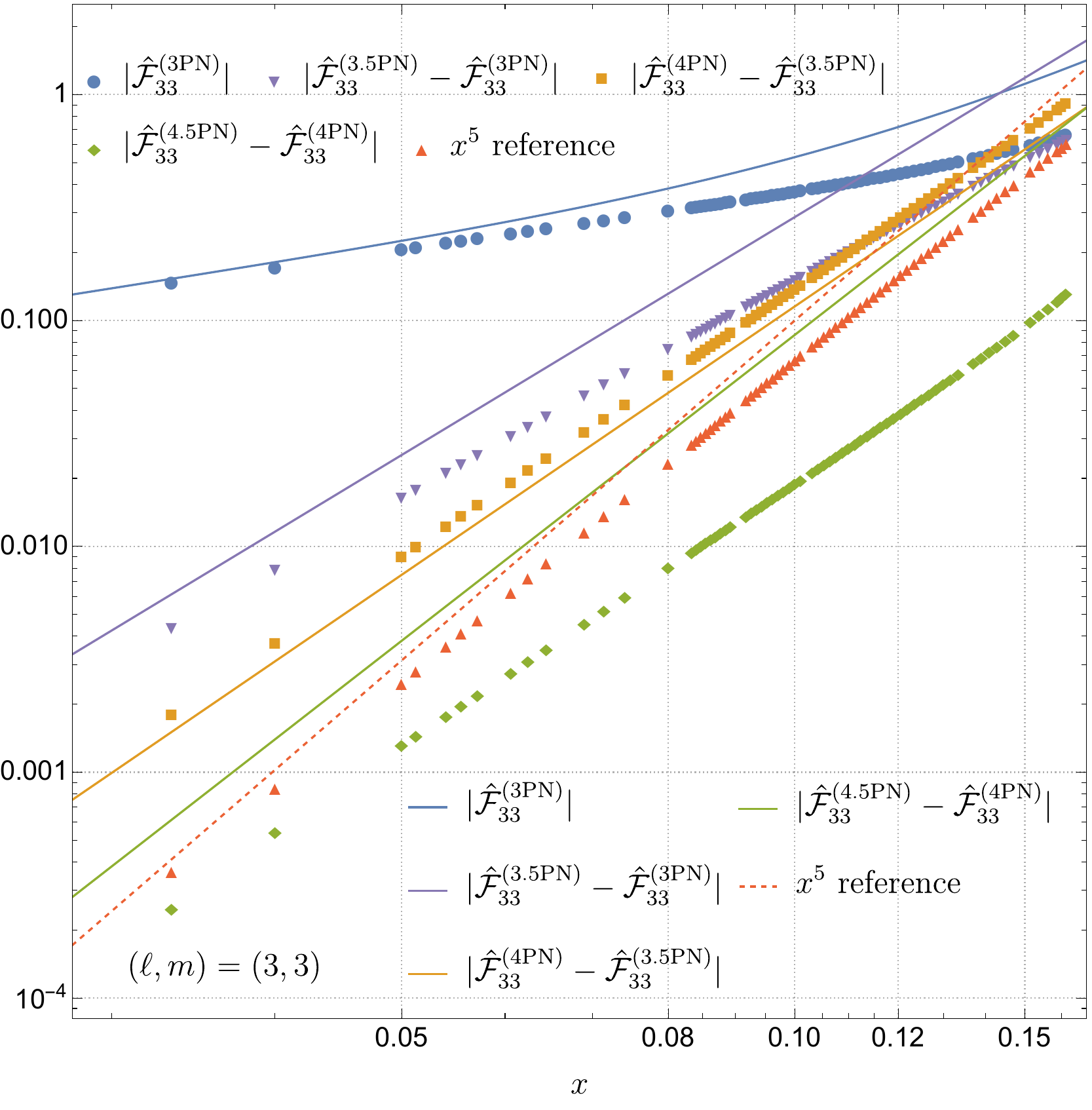}\hfill    \includegraphics[width=0.48\textwidth]{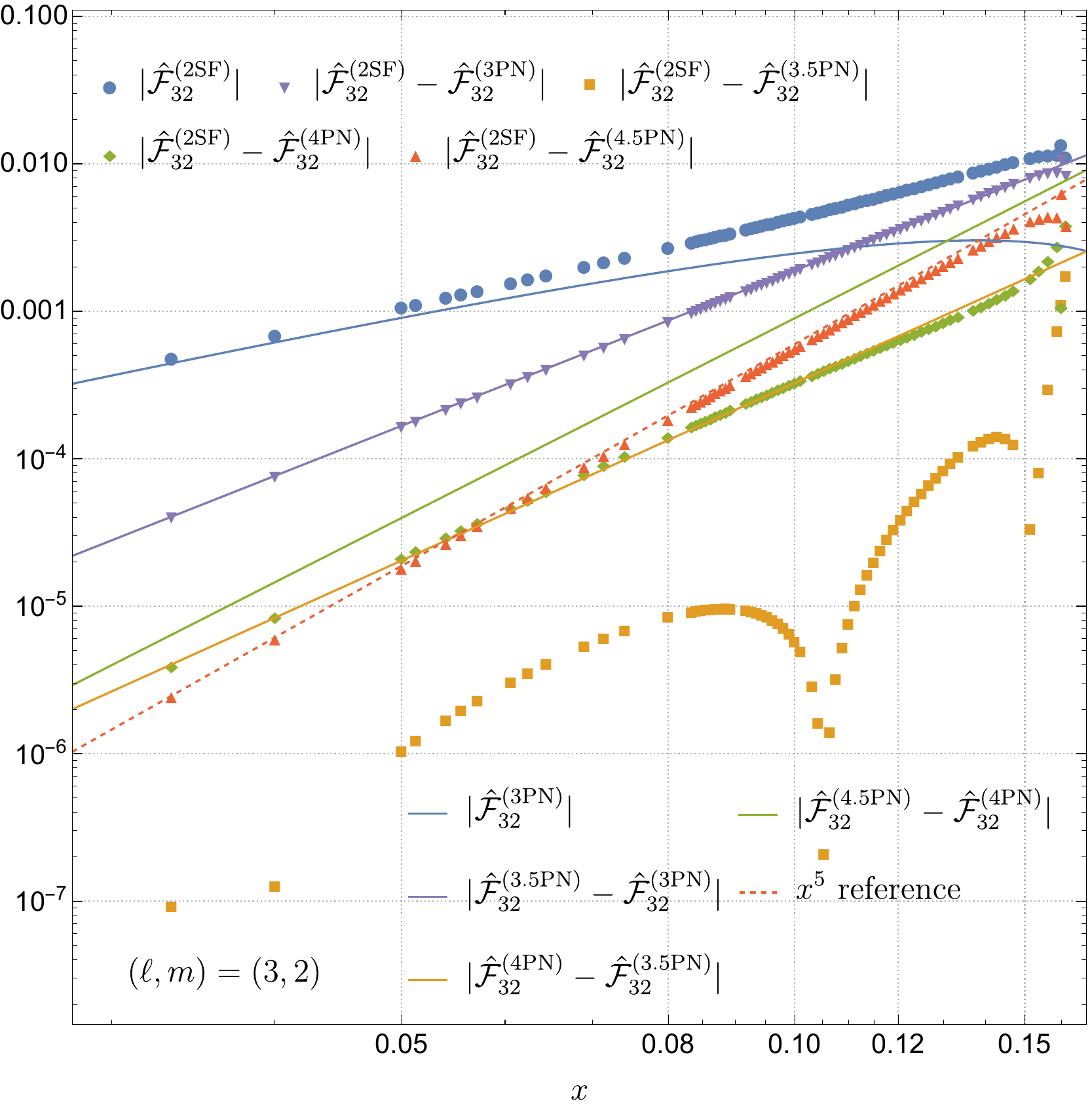}
	\caption{(\textit{Left panel}) 
    Detailed comparison of $\hat{\mathcal{F}}^{(2)}_{33}$, the $\mathcal{O}(\nu)$ coefficient of the $(3,3)$ mode of the (Newtonian-normalized) flux, computed from (i) numerical GSF and (ii) PN theory.
    The residuals between the GSF data and the various PN series (dotted lines) do not clearly follow the next term in the PN series (full lines) in this frequency range. However, the relative difference between the dotted lines and the corresponding full lines become smaller as one goes to smaller frequencies. This suggests that at smaller frequencies, one could credibly reach the asymptotic regime where the two curves overlap.  Finally, the residual with the 4.5PN term approaches a 5PN slope, as shown by the $\mathcal{O}(x^5)$ reference line. (\textit{Right panel}) Same as the left panel but for the (3,2) mode.
    The residual with the 3PN series (purple triangles) agrees perfectly with the 3.5PN term (purple line). The residual with the 3.5PN terms (orange squares) undergoes a passage through zero, then becomes noisy for $x < 0.05$, so it is not possible to claim agreement with the 4PN term (yellow line). Nonetheless, when going to the next order, we find that the residual with the 4PN series (green diamonds) agrees fairly well with the 4.5PN term (green line). Although the asymptotic regime is not yet reached, the relative error between the two curves decreases for smaller frequencies, and suggests better agreement deeper in the weak field regime. Finally, the residual with the 4.5PN term exhibits a very clear 5PN slope, as shown by the perfect fit with a $\mathcal{O}(x^5)$ reference line.\footnote{There should be also some logarithmic terms $\mathcal{O}(x^5 \ln x)$, but one cannot distinguish the slopes of $x^5$ and $x^5 \ln x$ on the scale of the plot.}
\label{fig:comparisonmodes_l3m3_l3m2}}
\end{figure*}

We present the results of our extensive comparison between the numerical 2SF data and the analytic 4.5PN flux given by Eq.~\eqref{eq:Flux}. 
In computing the total 2SF flux, we sum the $(\ell, m)$ modes up to $\ell = 6$
(the contribution from modes with higher $\ell$ is numerically small and does not substantially affect the comparison).
For the purpose of the GSF and PN comparison, we define the Newtonian-normalized flux as
\begin{align}\label{eq:normalizedflux}
    \hat{\mathcal{F}} \equiv \frac{\mathcal{F}}{\mathcal{F}_{\rm N}}\,,\qquad \hat{\mathcal{F}}_{\ell m} \equiv \frac{\mathcal{F}_{\ell m}}{\mathcal{F}_{\rm N}}\,,
\end{align}
where the Newtonian flux reads
\begin{align}
	\mathcal{F}_{\rm N} = \frac{32}{5}\nu^2 x^5\,.
\end{align}

In Fig.~\ref{fig:comparison_total_flux} we plot the GSF data for $\hat{\mathcal{F}}^{(2)} = \mathcal{F}^{(2)}/\mathcal{F}_\mathrm{N}$, i.e. corresponding to the terms in the normalized flux~\eqref{eq:normalizedflux} that are subleading in the mass ratio, together with the~PN predictions at various orders, up to 4.5PN. We find that both the 4PN and 4.5PN are numerically very close to the 2SF prediction until the relativistic regime at around $x\simeq 0.15$. For $x \lesssim 0.12$, the 4.5PN~series is in better agreement with the numerical data than the 4PN series; but around $x\simeq 0.15$, it is the opposite.  Lower-order PN approximations clearly agree less well. While the 3PN prediction is still rather good, the half-integer (odd-parity) 3.5PN is off. By contrast, 2.5PN is surprisingly close to the 2SF result, while the even-parity 2PN is the worst approximation in the series. We also note  that 1PN does relatively well. Here the PN expansion is in Taylor-expanded form, without resummation techniques applied.  

Figure~\ref{fig:detailed_comparison_total_flux} goes further in the comparison by checking the residuals between the GSF data and PN series.
We find that the residual obtained by subtracting the 3.5PN result from the numerical GSF data clearly follows the 4PN term.
When further subtracting the 4PN series from the numerical GSF data, we observe that the amplitude of the residual is subdominant (with respect to the 4PN term) and close to the 4.5PN term. However, over the range of frequencies where we have GSF data, we cannot ascertain the slope nor conclusively say that the residual is tending to the 4.5PN result.
Finally, after subtracting the 4.5PN term from the numerical GSF data, we again observe that the amplitude of the residual is subdominant but again, we cannot ascertain how the residual scales with $x$.
Our inability to conclusively determine the scaling of the 4PN and 4.5PN residual stems from the limited range of frequencies over which we have accurate numerical GSF data. 
As the PN series is an asymptotic expansion, we only expect agreement with the numerical data for sufficiently small values of $x$.
How small $x$ has to be depends on the PN residual being considered. In particular, the scaling of the residual will only become apparent for values of $x$ below which there are no more zero crossings. For example, in Fig.~\ref{fig:detailed_comparison_total_flux}, there is a zero crossing in the 4.5PN residual at $x \simeq 0.11$. These zero crossings can in fact occur at much smaller values of $x$ than we have access to; we give an illustration of this by studying equivalent plots for the \textit{first-order} flux in Appendix \ref{apdx:1SFcomp}.
Thus, firmly establishing the scaling of the 4PN and 4.5PN residuals would require accurate GSF data at much smaller values of $x$ than we currently have access to.
The level of this noise is consistent with the error in each $(\ell, m)$ being similar to the estimated error in the (2,2) mode as shown in Fig.~7~of~\cite{WPWMD21}.

Although the comparison with the total flux in Fig.~\ref{fig:detailed_comparison_total_flux} provides evidence of mutual consistency between the GSF and PN results, the precise scaling of the residual with 4PN and 4.5PN cannot be clearly established.
Fortunately, we can build further confidence in our results by making comparisons at the level of individual $(\ell,m)$ modes, because each PN order and each $(\ell,m)$ mode exhibits different zero crossings.
The comparison with the $(2,2)$ mode is shown in Fig.~\ref{fig:flux22}.
For this mode, the comparison with the 3.5PN term pushes the residual into the numerical noise of the GSF data at small values of~$x$.
There is also evidence that there must be a further zero crossing in the 4PN residual at smaller values of $x$ than we have access to. 
The comparison with the $(3,3)$ and $(3,2)$ modes is much less noisy, even down to the 4.5PN residuals, as shown in Fig.~\ref{fig:comparisonmodes_l3m3_l3m2}.
Similarly, good agreement is seen for many other modes -- see the figures in Appendix \ref{app:Flm-comparison}.
We note here that agreement between the $(\ell,m)$ modes is not guaranteed unless the waveforms are computed in the same asymptotic Bondi-Metzner-Sachs (BMS) frames~\cite{Boyle:2015nqa}.
At present, in second-order GSF calculations, this is not well understood and requires more work; we briefly discuss this further in Appendix \ref{app:Flm-comparison}.

\begin{figure*}[htp]
  \centering
  \includegraphics[width=0.48\textwidth]{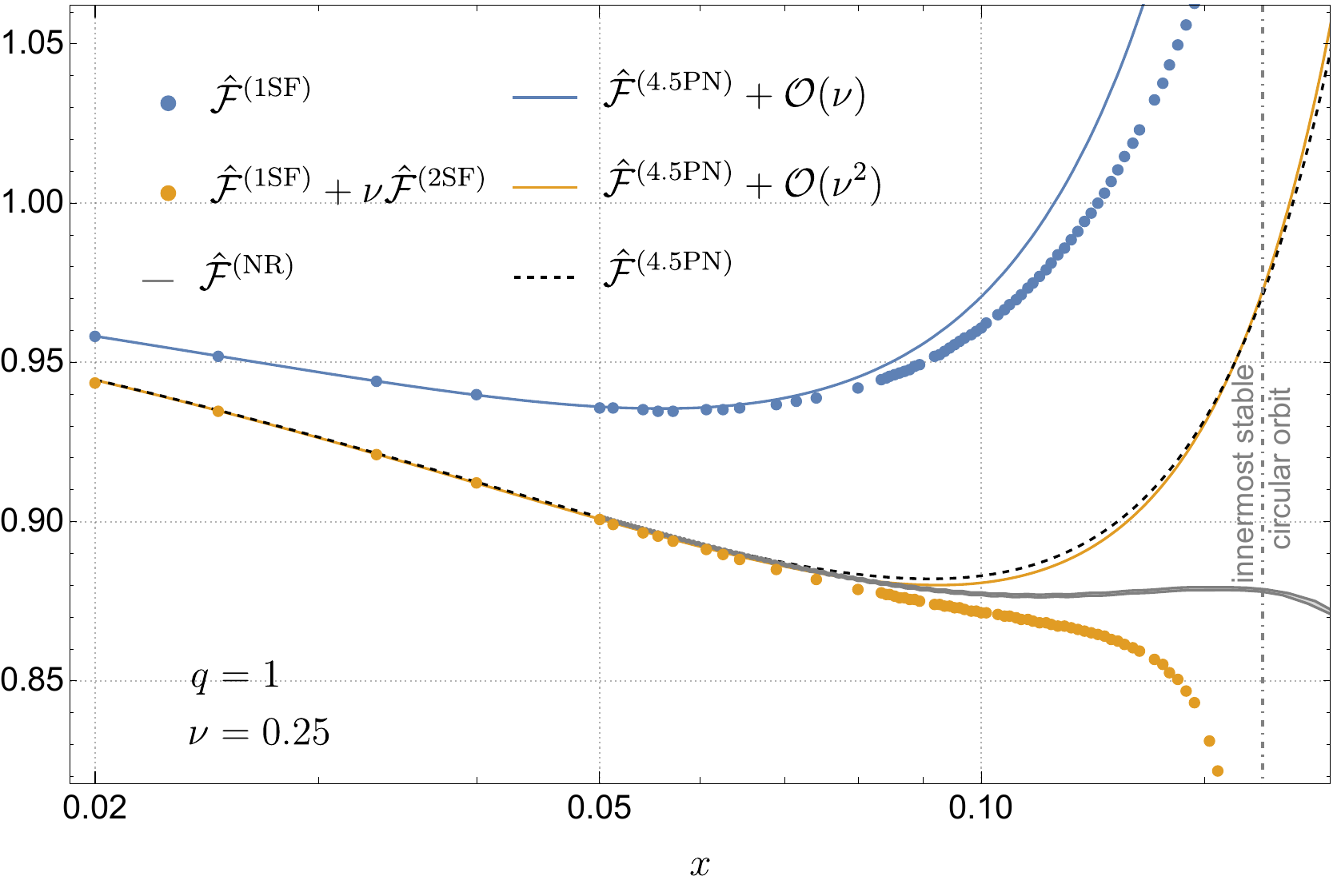}\hfill
  \includegraphics[width=0.48\textwidth]{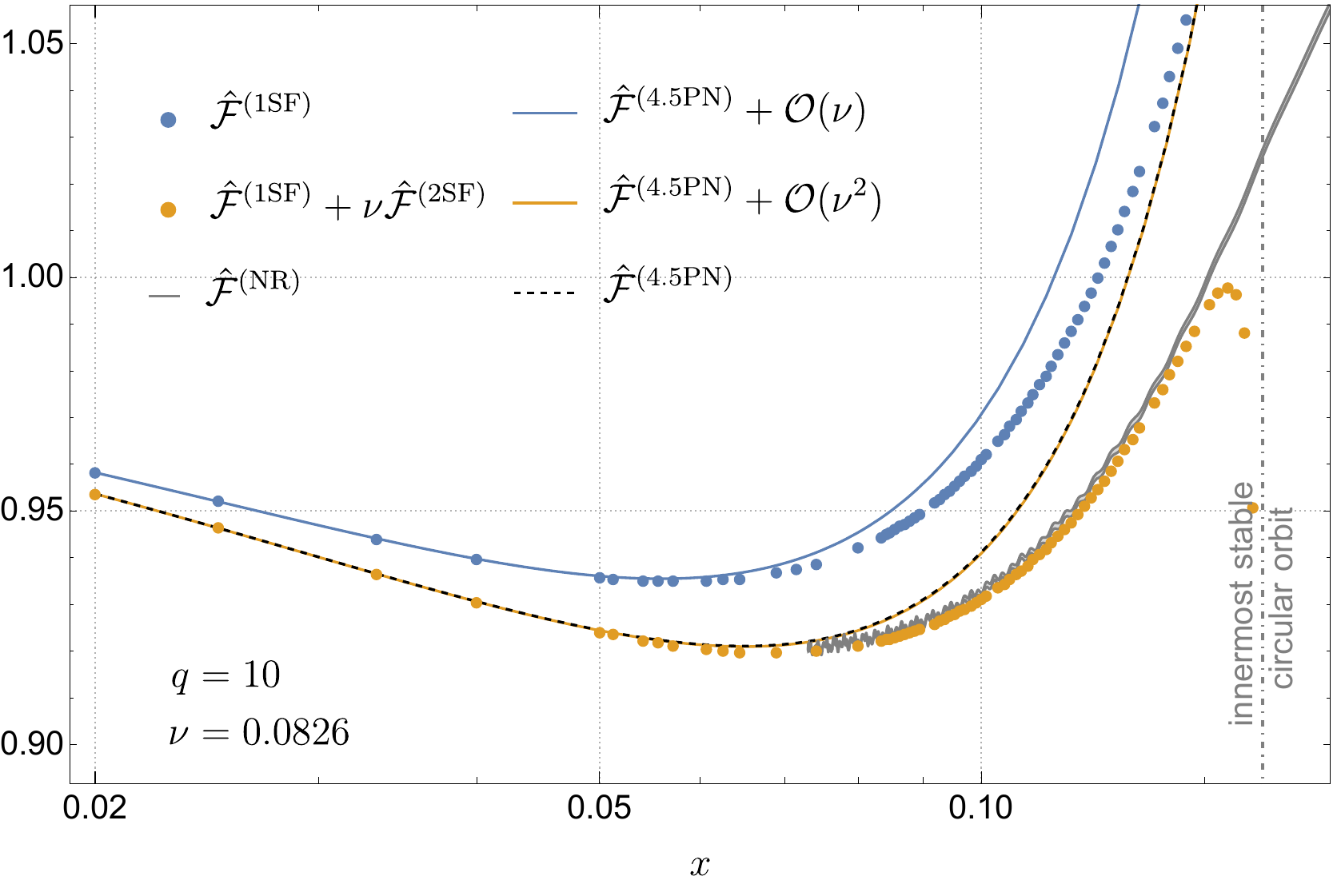}
  \caption{(\textit{Left Panel}) Comparison between the (Newtonian-normalized) GSF and PN total flux at different orders in the mass ratio expansion for $q=1$ ($\nu=0.25$). It is interesting to note that the 4.5PN expansion, including all orders in $\nu$ (black, dashed curve), is closely approximated by the PN expansion truncated at $\mathcal{O}(\nu^2)$ (yellow, solid curve), despite the equal mass ratio. The two gray curves show the flux computed from an NR simulation by the SXS collaboration~\cite{Boyle_2019}, specifically SXS:BBH:1132 \cite{SXS:BBH:1132}. The two gray curves are computed using, respectively, 3rd and 4th order polynomials to extrapolate the NR waveform from the edge of the computational domain to null infinity and thus the shaded gray region between them is an estimate on the numerical error in the simulation. (\textit{Right Panel}) The same as the left panel but for $q=10$ ($\nu=0.0826$). The NR simulation in this plot is SXS:BBH:1107~\cite{SXS:BBH:1107}. 
  }
\label{fig:comparison1SFplus2SF}
\end{figure*}

In Fig.~\ref{fig:comparison1SFplus2SF} we assess the accuracy and convergence of the SF fluxes when benchmarked against PN and numerical relativity (NR); this complements the analogous assessment of PN in Fig.~\ref{fig:comparison_total_flux}. Here we include the normalized total flux at 1SF [i.e., neglecting $\calO(\nu)$] and 2SF [neglecting $\calO(\nu^2)$] alongside 4.5PN and NR results for two mass ratios: $q=1$ (left panel) and $q=10$ (right panel). We see that for both mass ratios, the 2SF flux provides a marked improvement over the 1SF result. The SF expansion, moving from 1SF to 2SF, converges well toward the NR result, achieving nice agreement with NR for the large mass ratio $q=10$ throughout the strong field regime $x\gtrsim 0.07$ where the NR data is available. The SF expansion also converges nicely toward the PN result in the weak field, with the PN and 2SF results agreeing extremely well in the mildly relativistic regime, say $x\lesssim 0.07$, for both mass ratios. To further illustrate the convergence, we also include curves for the 4.5PN flux truncated at 1SF and at 2SF. The latter agrees extremely well with the full 4.5PN flux for all values of $x$, showing that higher-order corrections $\calO(\nu^2)$ are numerically small even at $q=1$. We note that for $q=1$, the 4.5PN flux remains more accurate than the 2SF flux, when measured against NR, even in the strong field up to $x\simeq 0.1$ (a very relativistic value by PN standards). Figure~\ref{fig:comparison1SFplus2SF} thus illustrates the importance of synergies between the PN, SF, and NR approaches to cover the whole range of scenarios of compact binary coalescences.

\section{Conclusion}

We have compared the results of two significant recent advances in perturbative techniques for GWs emitted by compact binary systems, focusing on the energy flux for quasicircular orbits: on the one hand, the analytical slow-motion weak-field post-Newtonian approximation, pushed to 4.5PN order beyond the Einstein quadrupole formula~\cite{BFHLT23a, BFHLT23b}; on the other hand, the numerical gravitational self-force approach developed to second order in the small mass ratio limit~\cite{WPWMD21,WPWMDT23}.

To 1SF level, the GSF flux has been derived analytically to high order in the PN expansion~\cite{TSasa94,TTS96,Fuj14PN,Fuj22PN}, and the PN coefficients are perfectly consistent up to 4.5PN order with direct PN calculations valid for arbitrary mass ratios. Note that at 1SF/0PA order, the flux at infinity for quasicircular inspirals is identical to the flux generated by a particle on a fixed background geodesic.

The new comparisons at the 2SF level reported in this paper are numerical, building\footnote{The second-order self-force data used in Ref.~\cite{WPWMD21} contained an error that was numerically small but had a significant impact on some of the PN comparisons in that paper. An erratum giving full details is currently in preparation. The same error affected an earlier version of this paper [\href{https://arxiv.org/abs/2407.00366v1}{arXiv:2407.00366v1}]. Using the incorrect data, we found good agreement at the level of the total flux, but stark disagreement at the level of individual modes. We interpreted this as mode-mixing, and suggested that it could arise from differences in the asymptotic frames used in the two calculations. Correcting this error in the data removed this apparent mode-mixing; we have therefore revised some of our conclusions.} on earlier comparisons with the 3.5PN flux in Ref.~\cite{WPWMD21}. The 2SF fluxes take into account the contribution coming from the 1SF deviation in the motion of the particle generating the GWs and flux at infinity, as well as quadratic nonlinearities in waveform generation and propagation.

Over the considered range of frequencies ($0.3< x <0.16$), we found that the 4PN and 4.5PN series agree much better with the numerical SF plot that any other lower-order PN truncation, see Fig. \ref{fig:comparison_total_flux}. We then studied the weak-field scaling of the residuals between the SF data and PN series at different orders, and found that the numerical data is compatible with the PN coefficients for the total flux, although we cannot conclusively demonstrate that the residual with the 4PN and 4.5PN results scale as expected over the range of frequencies for which we have GSF data. This is most likely due to the presence of zero crossings, which indicate that we have not yet reached the asymptotic regime where the comparison can be performed. Reaching this regime would require reducing the numerical noise in the 2SF data to levels significantly lower than what is currently achievable. To avoid any doubt, this does not mean that the numerical and PN fluxes disagree: one should keep in mind that near a zero crossing, the truncated PN series  happen to agree numerically with the GSF data \textit{better} that what is predicted by the asymptotic PN scaling. 
Finally, further comparisons at the level of the individual $(\ell,m)$ modes add further weight to the agreement, since the residuals for some $(\ell,m)$ modes do scale in the expected way -- see Figs.~\ref{fig:flux22}, \ref{fig:comparisonmodes_l3m3_l3m2}, \ref{fig:fluxl4}, and \ref{fig:comparisonmodes_l5m5_l6m6}.

Finally, note that a key aspect of our analysis is that invariant comparisons require precise knowledge of how the orbital frequency relates to the waveform frequency. Agreement between the 4.5PN and 2SF fluxes is only achieved when the fluxes are computed as functions of an invariant waveform frequency, eliminating the gauge/slicing dependence of the relationship between the orbit and the waveform~\cite{T25_Schott}.

\section{Acknowledgments}
We are indebted to François Larrouturou for initiating this work with us, and for interesting discussions at an early stage of the project. A.P. is grateful to Thibault Damour for collaboration on a related project that uncovered an error in the 2SF flux data. 
N.W. acknowledges support from a Royal Society -- Science Foundation Ireland University Research Fellowship. This publication has emanated from research conducted with the financial support of Science Foundation Ireland under grant number 22/RS-URF-R/3825. D.T. received support from the Czech Academy of Sciences under the grant number LQ100102101. A.P.~acknowledges the support of a Royal Society University Research Fellowship and the UKRI Frontier Research Grant GWModels, as selected by the ERC and funded by UKRI under the Horizon Europe Guarantee scheme [grant number EP/Y008251/1]. This work is supported by ERC grant EMRIWaveforms (DOI: \href{https://doi.org/10.3030/101200625}{10.3030/101200625}). This work makes use of the Black Hole Perturbation Toolkit.

\appendix

\section{PN expansions for individual mode contributions to the flux}
\label{app:Flm}

Some of the comparisons in this paper involve individual $(\ell,m)$ modes of the flux. As these were not given previously in the literature, we give them explicitly here. 
At 4PN order, the modes can be straightforwardly obtained by inserting the $(2,2)$ mode amplitude~\eqref{eq:h22} and Eqs.~(3.4) of~\cite{H23} for the other modes, into Eq.~\eqref{eq:defFlm}. Moreover, we can extend this result at 4.5PN order with the sole knowledge of the hereditary part of the first time derivative of the radiative moments~\cite{MBF16}. Using (2.6)--(2.7) of~\cite{BFIS08} it becomes
\begin{align}
	\mathcal{F}_{\ell m} &= \frac{1}{16\pi}\Bigl[\vert\dot{\dU}_{\ell m}\vert^2 + \vert\dot{\dV}_{\ell m}\vert^2 \Bigr]\,,
\end{align}
where 
\begin{subequations}\label{UV}\begin{align}
		\dU_{\ell m} &= \frac{4}{\ell!}\,\sqrt{\tfrac{(\ell+1)(\ell+2)}{2\ell(\ell-1)}}
		\,\alpha_L^{\ell m}\,\dU_L\,,\\ \dV_{\ell m} &=
		-\frac{8}{\ell!}\,\sqrt{\tfrac{\ell(\ell+2)}{2(\ell+1)(\ell-1)}}
		\,\alpha_L^{\ell m}\,\dV_L\,.
\end{align}\end{subequations}
We have defined the STF tensorial coefficient \mbox{$\alpha_L^{\ell m}=\int \dd\Omega_2\,\hat{n}_L(\theta,\phi) {Y}_{\ell m}^*(\theta,\phi)$}, see e.g. (4.7) of~\cite{HFB21} for an explicit expression.

The explicit 4.5PN expressions of the decomposition of the total flux are given in Eq.~\eqref{eq:FlmExpressions} hereafter; we have verified that these expressions agree perfectly with Eqs.~(A.3-27) of~\cite{TTS96} at first order in the mass ratio. We present the results in terms of the normalized fluxes $\hat{\mathcal{F}}_{\ell m}$ defined in Eq.~\eqref{eq:normalizedflux}.
We also provide supplemental material containing the modes of the normalized fluxes in a digital format \cite{SupplementalMaterial}.
\begin{widetext}
\begin{subequations}\label{eq:FlmExpressions}
\begin{align}
    \hat{\mathcal{F}}_{22} &= 1 + \left(- \frac{107}{21}+ \frac{55}{21}\nu\right)x+4\pi x^{3/2} + \left(\frac{4784}{1323}-\frac{87691}{5292}\nu+\frac{5851}{1323}\nu^2\right)x^2  + \left(- \frac{428}{21}+ \frac{178}{21}\nu\right)\pi x^{5/2} \nn\\
    & \ \ \,\quad+\left(\frac{99210071}{1091475}- \frac{1712}{105}\gamma_E + \frac{16}{3}\pi^2 - \frac{856}{105}\ln(16x) + \nu\left[\frac{1650941}{349272}+ \frac{41}{48} \pi^2\right] - \frac{669017}{19404}\nu^2 + \frac{255110}{43659} \nu^3\right) x^3\nn\\
    &\ \ \,\quad + \left(\frac{19136}{1323}- \frac{144449}{2646}\nu + \frac{33389}{2646}\nu^2 \right)\pi x^{7/2} \nn\\
   &\ \ \,\quad +\Bigg(- \frac{27956920577}{81265275} + \frac{183184}{2205}\gamma_\mathrm{E}- \frac{1712}{63}\pi^2 + \frac{91592}{2205}\ln(16x) \nn\\
   &\ \ \,\quad \qquad + \nu\left[ \frac{440671815511}{57210753600}+ \frac{21152}{441}\gamma_\mathrm{E}+\frac{87757}{16128}\pi^2 + \frac{10576}{441}\ln(16x)\right] \nn\\
    &\ \ \,\quad \qquad+ \nu^2\left[\frac{488191373}{4643730}- \frac{205}{224}\pi^2\right] - \frac{249670210}{5108103}\nu^3 + \frac{4763492}{729729}\nu^4 \Bigg)x^4 \nn\\
    & \ \ \,\quad +\Bigg(\frac{396840284}{1091475}- \frac{6848}{105}\gamma_\mathrm{E} - \frac{3424}{105}\ln(16x) + \nu \left[\frac{1882981}{174636}+ \frac{41}{12}\pi^2\right] - \frac{885179}{8316}\nu^2 + \frac{2627879}{174636}\nu^3 \Bigg)\pi x^{9/2}+ \mathcal{O}(x^5) \label{eq:h22PN}\\
    \hat{\mathcal{F}}_{21} &= \left(\frac{1}{36}- \frac{\nu}{9}\right)x+\left(- \frac{17}{504}+\frac{11}{63}\nu - \frac{10}{63}\nu^2\right)x^2 + \left(\frac{1}{18} - \frac{2}{9} \nu\right)\pi x^{5/2} \nn\\
    & \quad+ \left(- \frac{2215}{254016}- \frac{13567}{63504}\nu + \frac{65687}{63504} \nu^2  - \frac{853}{5292} \nu^3\right)x^3 + \left(- \frac{17}{252} + \frac{9}{28}\nu  - \frac{13}{63} \nu^2\right) \pi x^{7/2} \nn\\
    &\quad +\Bigg(\frac{15707221}{26195400} - \frac{107}{945}\gamma_\mathrm{E}+ \frac{\pi^2}{27}  - \frac{107}{1890} \ln(4x) + \nu\left[- \frac{17451737}{3742200}+ \frac{428}{945}\gamma_\mathrm{E}-\frac{409}{6912}\pi^2 + \frac{214}{945}\ln(4x)\right] \nn\\
    &\ \ \,\quad \quad+ \nu^2\left[\frac{74234059}{8382528}- \frac{205}{576}\pi^2\right] + \frac{965}{1134}\nu^3 - \frac{70787}{523908}\nu^4\Bigg)x^4 \nn\\
    &\quad +\Bigg(-\frac{2215}{127008}  - \frac{23669}{63504} \nu + \frac{115141}{63504}\nu^2 - \frac{305}{1764}\nu^3 \Bigg)\pi x^{9/2}+ \mathcal{O}(x^5) \\
    \hat{\mathcal{F}}_{33} &= \left(\frac{1215}{896}- \frac{1215}{224}\nu\right)x+\left(- \frac{1215}{112}+\frac{10935}{224}\nu - \frac{1215}{56}\nu^2\right)x^2 + \left(\frac{3645}{448} - \frac{3645}{112} \nu\right)\pi x^{5/2} \nn\\
    & \quad+ \left( \frac{243729}{9856}- \frac{92907}{616}\nu + \frac{2171691}{9856} \nu^2  - \frac{17901}{352} \nu^3\right)x^3 + \left(- \frac{3645}{56} + \frac{258795}{896}\nu  - \frac{3645}{32} \nu^2\right) \pi x^{7/2} \nn\\
    &\quad +\Bigg(\frac{25037019729}{125565440} - \frac{47385}{1568}\gamma_\mathrm{E}+ \frac{3645}{224}\pi^2  - \frac{47385}{3136} \ln(36x) \nn\\
    & \qquad\quad+ \nu\left[- \frac{84912893811}{125565440}+ \frac{47385}{392}\gamma_\mathrm{E}-\frac{1816425}{28672}\pi^2 + \frac{47385}{784}\ln(36x)\right] \nn\\
    &\qquad\quad+ \nu^2\left[-\frac{45782091}{73216}- \frac{49815}{7168}\pi^2\right] + \frac{149211477}{256256}\nu^3 - \frac{5737635}{64064}\nu^4\Bigg)x^4 \nn\\
    &\quad +\Bigg(\frac{731187}{4928}  - \frac{34272963}{39424} \nu + \frac{45879129}{39424}\nu^2 - \frac{2379213}{9856}\nu^3 \Bigg)\pi x^{9/2}+ \mathcal{O}(x^5) \label{eq:h33PN}\\
    \hat{\mathcal{F}}_{32} &= \left( \frac{5}{63}-\frac{10}{21}\nu + \frac{5}{7}\nu^2\right)x^2 + \left( -\frac{193}{567}+ \frac{1304}{567}\nu - \frac{2540}{567} \nu^2  + \frac{365}{189} \nu^3\right)x^3 + \left(\frac{20}{63} - \frac{40}{21}\nu  + \frac{20}{7} \nu^2\right) \pi x^{7/2} \nn\\
    &\quad +\Bigg(\frac{86111}{280665} - \frac{183244}{56133}\nu + \frac{709004}{56133}\nu^2 - \frac{4009259}{224532}\nu^3 + \frac{181429}{56133} \nu^4\Bigg)x^4 \nn\\
    &\quad +\Bigg(-\frac{772}{567}  + \frac{5126}{567} \nu - \frac{9620}{567}\nu^2 + \frac{170}{27}\nu^3 \Bigg)\pi x^{9/2}+ \mathcal{O}(x^5) \label{eq:h32PN}\\
    \hat{\mathcal{F}}_{31} &= \left(\frac{1}{8064}- \frac{\nu}{2016}\right)x+\left(- \frac{1}{1512}+\frac{5}{2016}\nu + \frac{\nu^2}{1512}\right)x^2 + \left(\frac{1}{4032} - \frac{\nu}{1008} \right)\pi x^{5/2} \nn\\
    & \quad+ \left( \frac{437}{266112}- \frac{1291}{199584}\nu - \frac{523}{798336} \nu^2  + \frac{29}{28512} \nu^3\right)x^3 + \left(- \frac{1}{756} + \frac{13}{2688}\nu  + \frac{11}{6048} \nu^2\right) \pi x^{7/2} \nn\\
    &\quad +\Bigg(-\frac{1137077}{50854003200} - \frac{13}{42336}\gamma_\mathrm{E}+ \frac{\pi^2}{6048}  - \frac{13}{84672} \ln(4x) \nn\\
    & \qquad\quad+ \nu\left[- \frac{117030737}{50854003200}+ \frac{13}{10584}\gamma_\mathrm{E}-\frac{389}{774144}\pi^2 + \frac{13}{21168}\ln(4x)\right] \nn\\
    &\qquad\quad+ \nu^2\left[\frac{8245387}{622702080}- \frac{41}{64512}\pi^2\right] - \frac{920879}{62270208}\nu^3 + \frac{4681}{15567552}\nu^4\Bigg)x^4 \nn\\
    &\quad +\Bigg(\frac{437}{133056}  - \frac{38903}{3193344} \nu - \frac{13259}{3193344}\nu^2 + \frac{1063}{798336}\nu^3 \Bigg)\pi x^{9/2}+ \mathcal{O}(x^5) \\
     \hat{\mathcal{F}}_{44} &= \left( \frac{1280}{567}-\frac{2560}{189}\nu + \frac{1280}{63}\nu^2\right)x^2 + \left( -\frac{151808}{6237}+ \frac{2995712}{18711}\nu - \frac{1853440}{6237} \nu^2  + \frac{32000}{297} \nu^3\right)x^3 \nn\\
    &\quad+ \left(\frac{10240}{567} - \frac{20480}{189}\nu  + \frac{10240}{63} \nu^2\right) \pi x^{7/2} \nn\\
    &\quad +\Bigg(\frac{560069632}{6243237} - \frac{66822586112}{93648555}\nu + \frac{24511650304}{13378365}\nu^2 - \frac{158797120}{99099}\nu^3 + \frac{95524096}{297297} \nu^4\Bigg)x^4 \nn\\
    &\quad +\Bigg(-\frac{1214464}{6237}  + \frac{23796736}{18711} \nu - \frac{14489600}{6237}\nu^2 + \frac{1623040}{2079}\nu^3 \Bigg)\pi x^{9/2}+ \mathcal{O}(x^5) \label{eq:h44PN}\\
    \hat{\mathcal{F}}_{43} &=  \left( \frac{729}{4480}- \frac{729}{560}\nu + \frac{729}{224} \nu^2  - \frac{729}{280} \nu^3\right)x^3 +\Bigg(-\frac{28431}{24640} + \frac{198045}{19712}\nu  - \frac{288441}{9856} \nu^2 + \frac{20169}{616}\nu^3 - \frac{31833}{3080} \nu^4\Bigg)x^4 \nn\\
    &\quad +\Bigg(\frac{2187}{2240}  - \frac{2187}{280} \nu + \frac{2187}{112}\nu^2 - \frac{2187}{140}\nu^3 \Bigg)\pi x^{9/2}+ \mathcal{O}(x^5) \\
     \hat{\mathcal{F}}_{42} &= \left( \frac{5}{3969}-\frac{10}{1323}\nu + \frac{5}{441}\nu^2\right)x^2 + \left( -\frac{437}{43659}+ \frac{7958}{130977}\nu - \frac{4120}{43659} \nu^2  + \frac{95}{14553} \nu^3\right)x^3 \nn\\
    &\quad+ \left(\frac{20}{3969} - \frac{40}{1323}\nu  + \frac{20}{441} \nu^2\right) \pi x^{7/2} \nn\\
    &\quad +\Bigg(\frac{7199152}{218513295} - \frac{140762423}{655539885}\nu + \frac{37048126}{93648555}\nu^2 - \frac{3504901}{24972948}\nu^3 - \frac{3097}{297297} \nu^4\Bigg)x^4 \nn\\
    &\quad +\Bigg(-\frac{1748}{43659}  + \frac{31502}{130977} \nu - \frac{2260}{6237}\nu^2 + \frac{50}{14553}\nu^3 \Bigg)\pi x^{9/2}+ \mathcal{O}(x^5) \\
    \hat{\mathcal{F}}_{41} &=  \left( \frac{1}{282240}- \frac{\nu}{35280} + \frac{\nu^2}{14112}  - \frac{\nu^3}{17640} \right)x^3 +\Bigg(-\frac{101}{4656960} + \frac{229}{1241856}\nu  - \frac{107}{206976} \nu^2 + \frac{\nu^3}{1848} - \frac{83}{582120} \nu^4\Bigg)x^4 \nn\\
    &\quad +\Bigg(\frac{1}{141120}  - \frac{\nu}{17640} + \frac{\nu^2}{7056} - \frac{\nu^3}{8820} \Bigg)\pi x^{9/2}+ \mathcal{O}(x^5) \label{eq:h41PN}\\
    \hat{\mathcal{F}}_{55} &=  \left( \frac{9765625}{2433024}- \frac{9765625}{304128}\nu + \frac{48828125}{608256} \nu^2  - \frac{9765625}{152064} \nu^3\right)x^3  \nn\\
    &\quad+\Bigg(-\frac{2568359375}{47443968} + \frac{1005859375}{2156544}\nu  - \frac{7919921875}{5930496} \nu^2 + \frac{48828125}{33696}\nu^3 - \frac{9765625}{23166} \nu^4\Bigg)x^4 \nn\\
    &\quad +\Bigg(\frac{48828125}{1216512}  - \frac{48828125}{152064} \nu + \frac{244140625}{304128}\nu^2 - \frac{48828125}{76032}\nu^3 \Bigg)\pi x^{9/2}+ \mathcal{O}(x^5) \label{eq:h55PN}\\
    \hat{\mathcal{F}}_{54} &=  \Bigg(\frac{4096}{13365} - \frac{8192}{2673}\nu  + \frac{28672}{2673} \nu^2 - \frac{40960}{2673}\nu^3 + \frac{20480}{2673} \nu^4\Bigg)x^4 + \mathcal{O}(x^5)\\
    \hat{\mathcal{F}}_{53} &=  \left( \frac{2187}{450560}- \frac{2187}{56320}\nu + \frac{2187}{22528} \nu^2  - \frac{2187}{28160} \nu^3\right)x^3  \nn\\
    &\quad+\Bigg(-\frac{150903}{2928640} + \frac{621837}{1464320}\nu  - \frac{206307}{183040} \nu^2 + \frac{38637}{36608}\nu^3 - \frac{729}{4160} \nu^4\Bigg)x^4 \nn\\
    &\quad +\Bigg(\frac{6561}{225280}  - \frac{6561}{28160} \nu + \frac{6561}{11264}\nu^2 - \frac{6561}{14080}\nu^3 \Bigg)\pi x^{9/2}+ \mathcal{O}(x^5) \\
    \hat{\mathcal{F}}_{52} &=  \Bigg(\frac{4}{40095} - \frac{8}{8019}\nu  + \frac{28}{8019} \nu^2 - \frac{40}{8019}\nu^3 + \frac{20}{8019} \nu^4\Bigg)x^4 + \mathcal{O}(x^5)\\
    \hat{\mathcal{F}}_{51} &=  \left( \frac{1}{127733760}- \frac{\nu}{15966720} + \frac{\nu^2}{6386688}   - \frac{\nu^3}{7983360} \right)x^3  \nn\\
    &\quad+\Bigg(-\frac{179}{2490808320} + \frac{713}{1245404160}\nu  - \frac{887}{622702080} \nu^2 + \frac{71}{62270208}\nu^3 - \frac{\nu^4}{77837760}\Bigg)x^4 \nn\\
    &\quad +\Bigg(\frac{1}{63866880}  - \frac{\nu}{7983360}  + \frac{\nu^2}{3193344} - \frac{\nu^3}{3991680} \Bigg)\pi x^{9/2}+ \mathcal{O}(x^5) \\
    \hat{\mathcal{F}}_{66} &=  \Bigg(\frac{26244}{3575} - \frac{52488}{715}\nu  + \frac{183708}{715} \nu^2 - \frac{52488}{143}\nu^3 + \frac{26244}{143} \nu^4\Bigg)x^4 + \mathcal{O}(x^5) \label{eq:h66PN}\\
    \hat{\mathcal{F}}_{64} &=  \Bigg(\frac{131072}{9555975} - \frac{262144}{1911195}\nu  + \frac{917504}{1911195} \nu^2 - \frac{262144}{382239}\nu^3 + \frac{131072}{382239} \nu^4\Bigg)x^4 + \mathcal{O}(x^5)\\
    \hat{\mathcal{F}}_{62} &=  \Bigg(\frac{4}{5733585} - \frac{8}{1146717}\nu  + \frac{28}{1146717} \nu^2 - \frac{40}{1146717}\nu^3 + \frac{20}{1146717} \nu^4\Bigg)x^4 + \mathcal{O}(x^5).
    \end{align}
    \end{subequations}
\end{widetext}

\section{Comparison of individual mode contributions to the flux}
\label{app:Flm-comparison}

Most of the comparisons in the body of the paper are for the total flux to infinity, summed over $(\ell, m)$ mode contributions. However, additional information about the PN and SF waveform amplitudes can be gleaned by comparing the individual $(\ell, m)$ contributions.

\begin{figure*}[htb!]
	\includegraphics[width=0.48\textwidth]{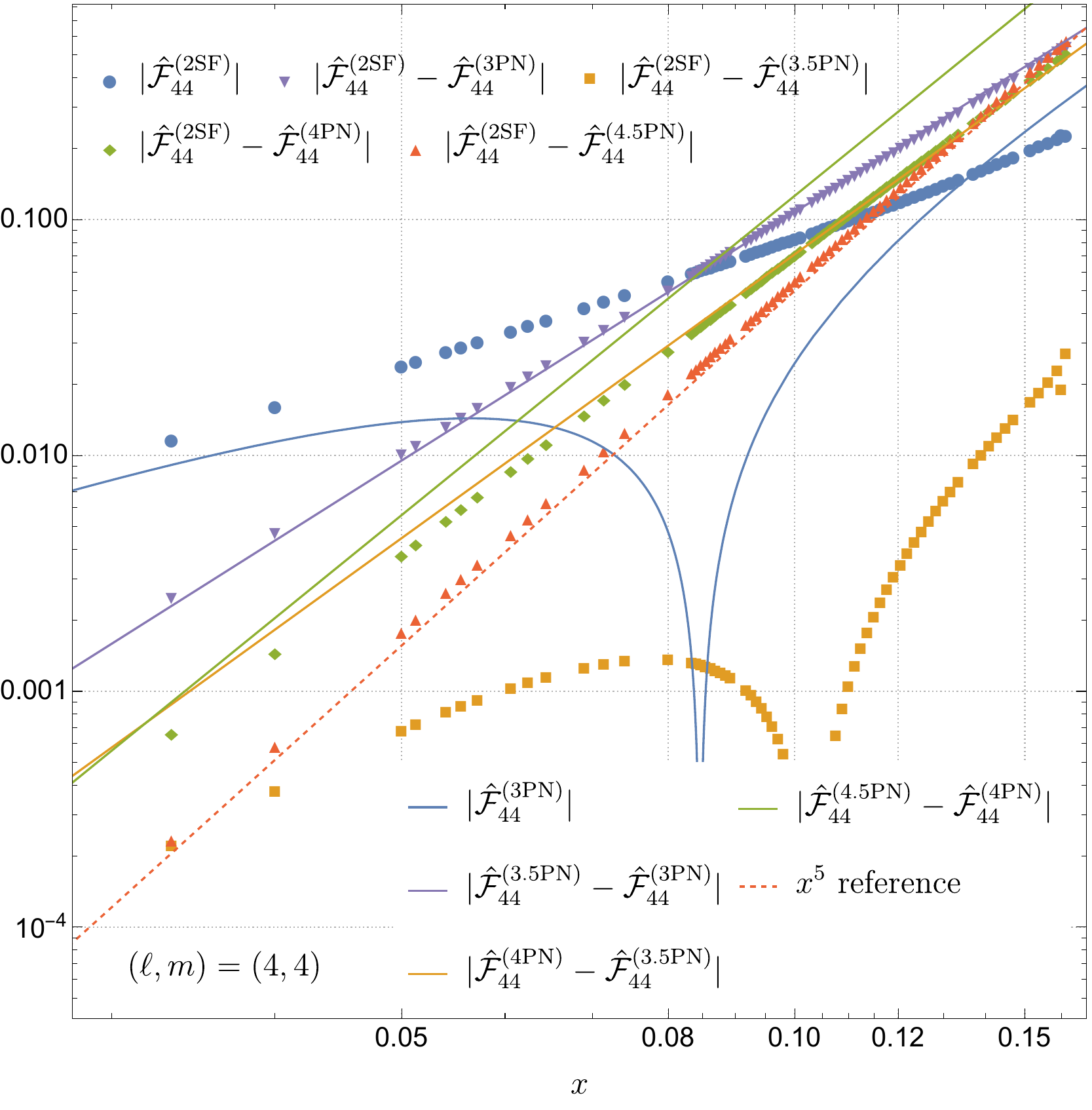}\hfill    \includegraphics[width=0.48\textwidth]{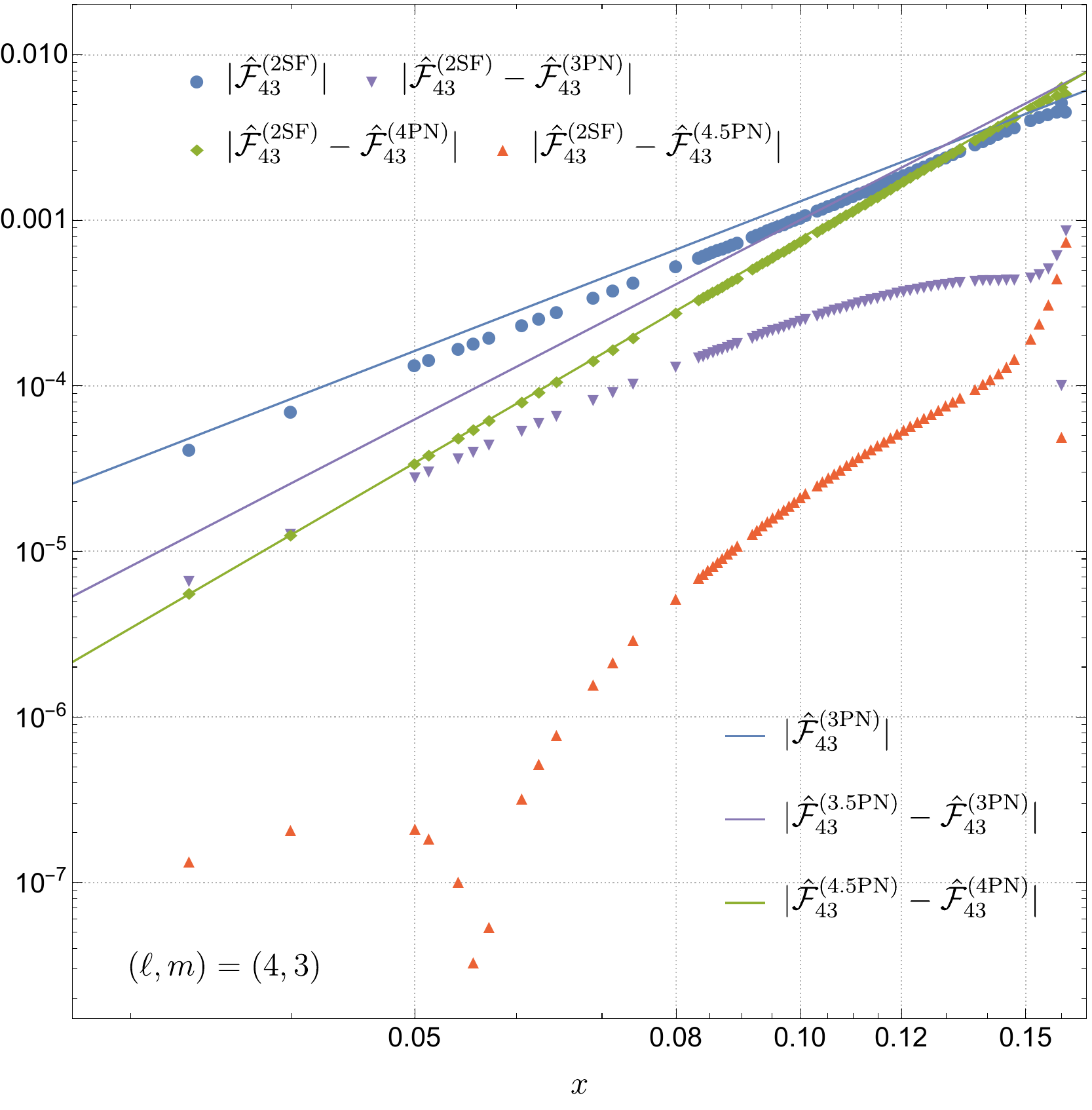} \\
    \includegraphics[width=0.48\textwidth]{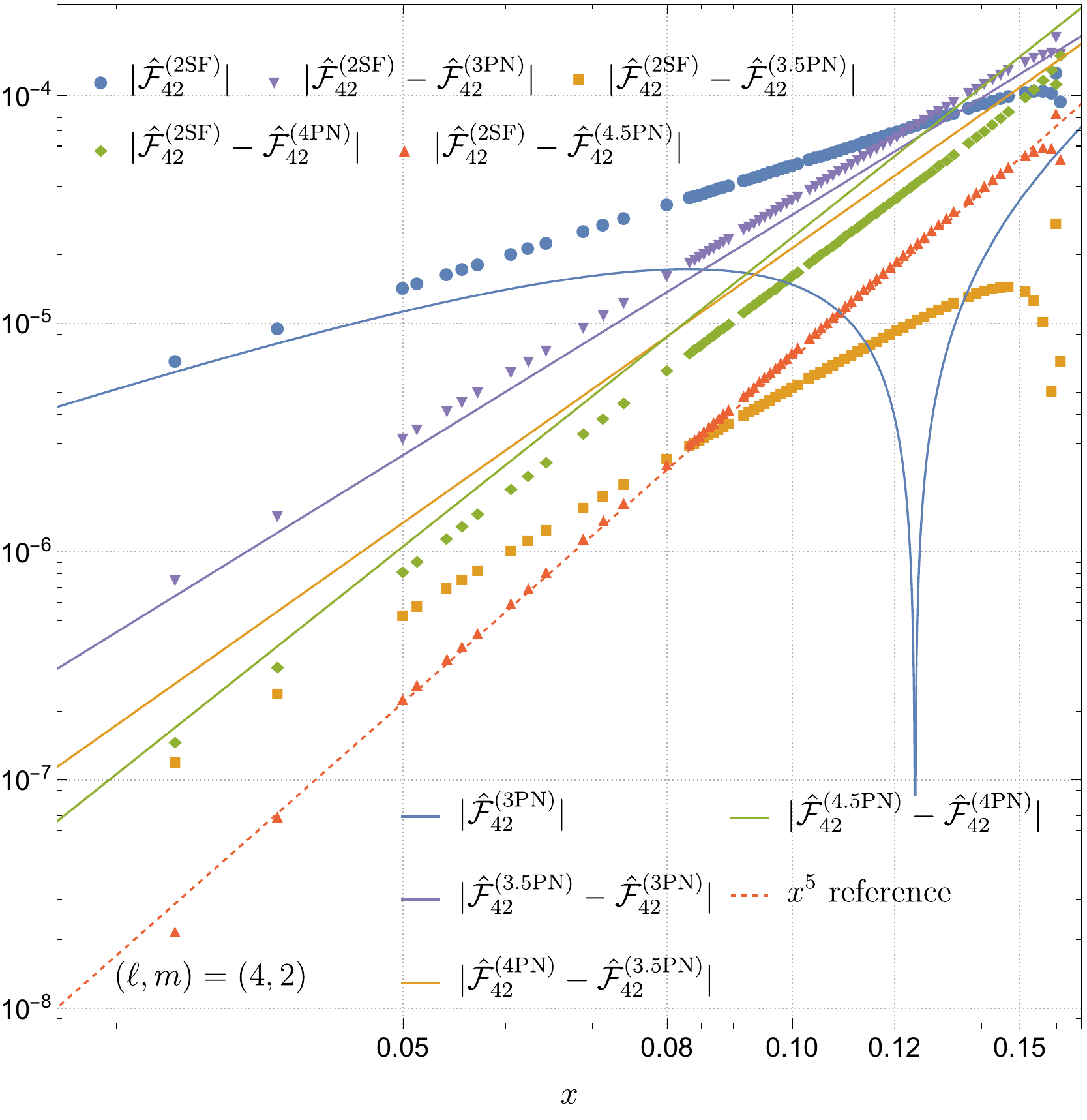}\hfill    \includegraphics[width=0.48\textwidth]{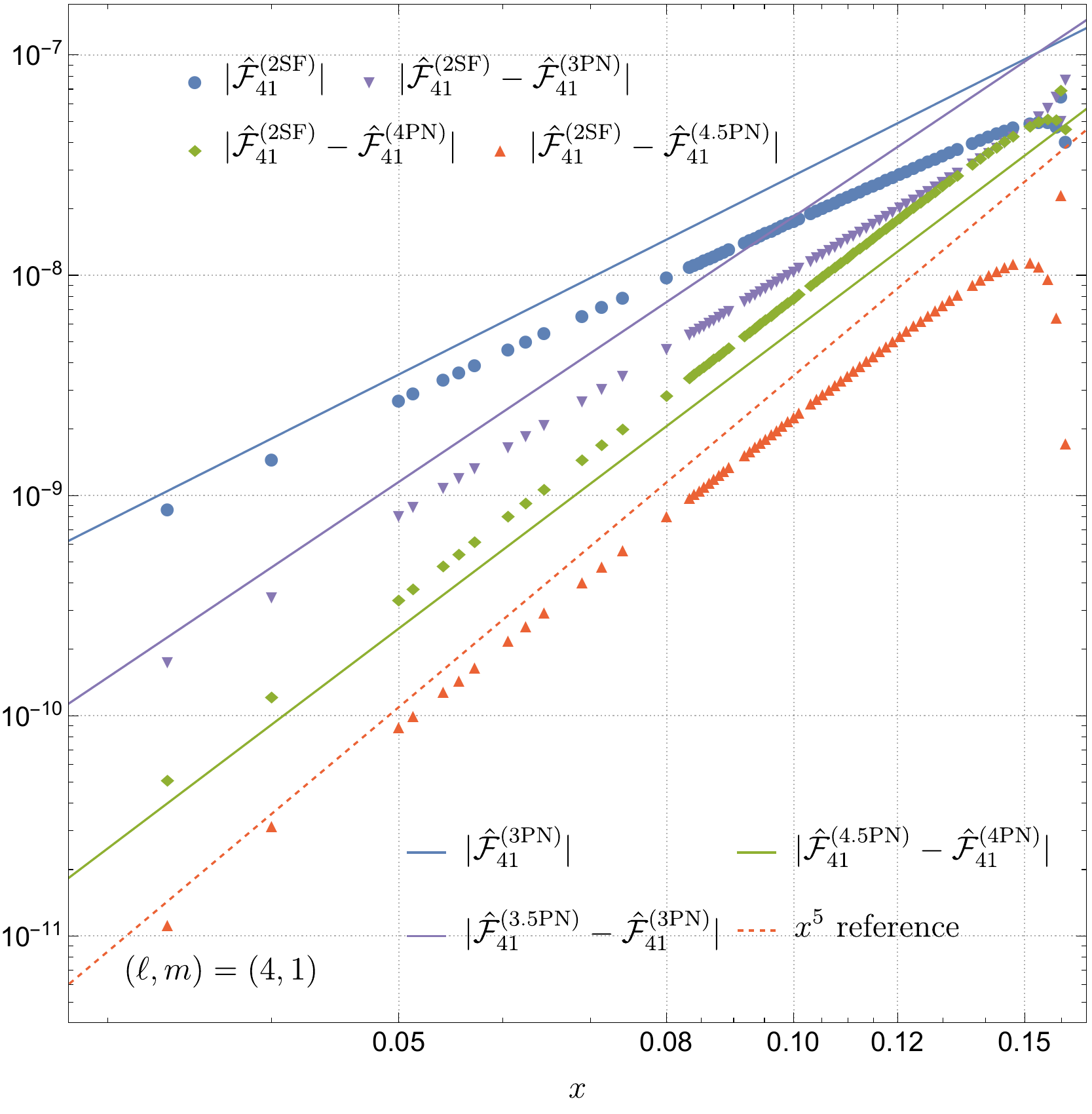} 
	\caption{
    Detailed comparison of $\hat{\mathcal{F}}^{(2)}_{4m}$ for $m\in\{4,3,2,1\}$, namely the $\mathcal{O}(\nu)$ coefficients of the $\ell=4$ modes of the (Newtonian-normalized) flux, computed from (i) numerical GSF and (ii) PN theory.
    For the $(4,4)$, $(4,2)$ and $(4,1)$ modes, the residual after subtracting the 4.5PN series exhibit a clear 5PN slope, as shown by the dashed (red) $x^5$ curve. In the comparison for the $(4,3)$ mode, there is a zero-crossing around $x\simeq0.055$ and we do not have GSF data for sufficiently small values of $x$ to see if the residual settles down to the expected fall-off rate. 
	}\label{fig:fluxl4}
\end{figure*}

\begin{figure*}[htb!]
	\includegraphics[width=0.48\textwidth]{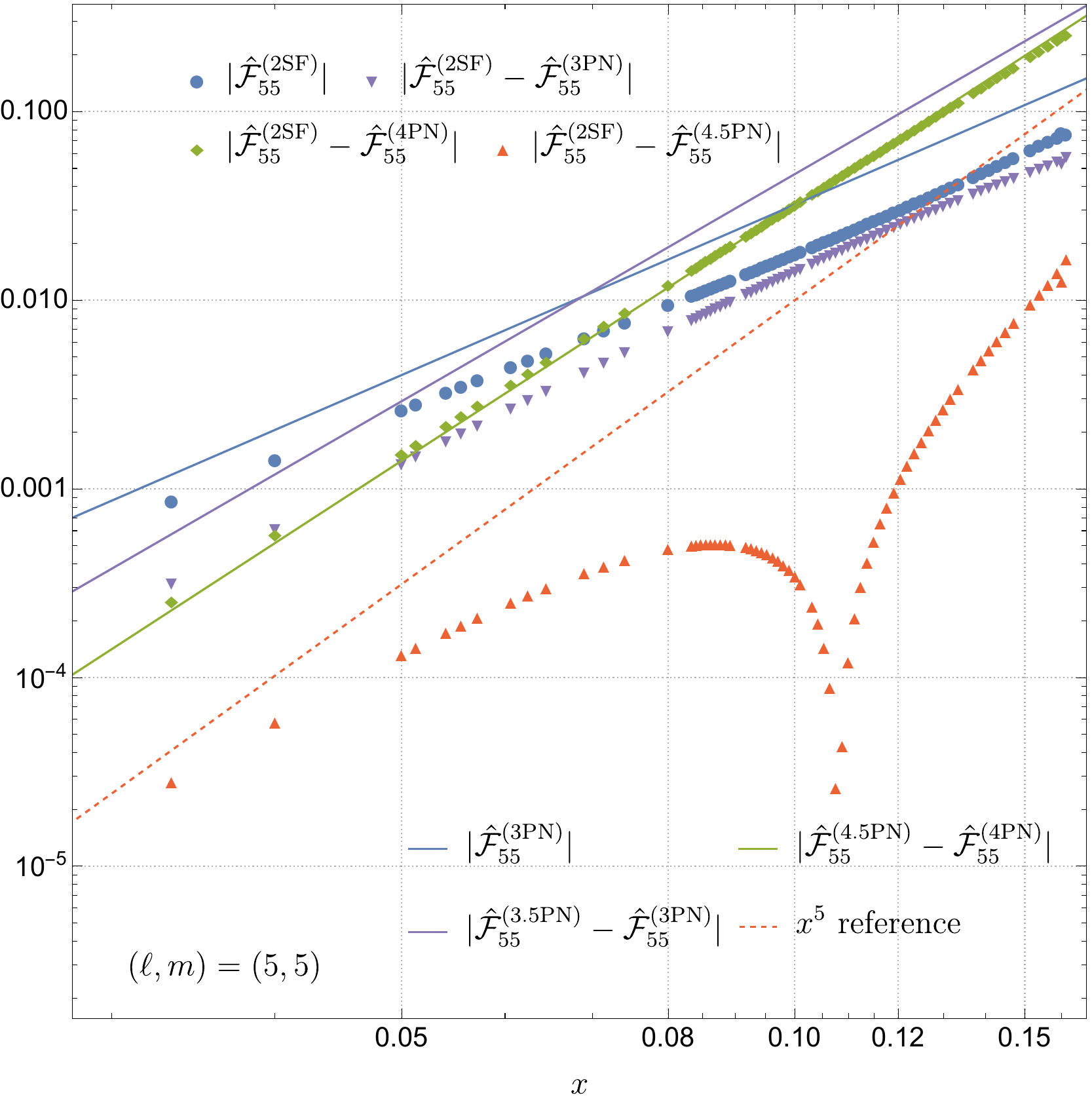}\hfill    \includegraphics[width=0.48\textwidth]{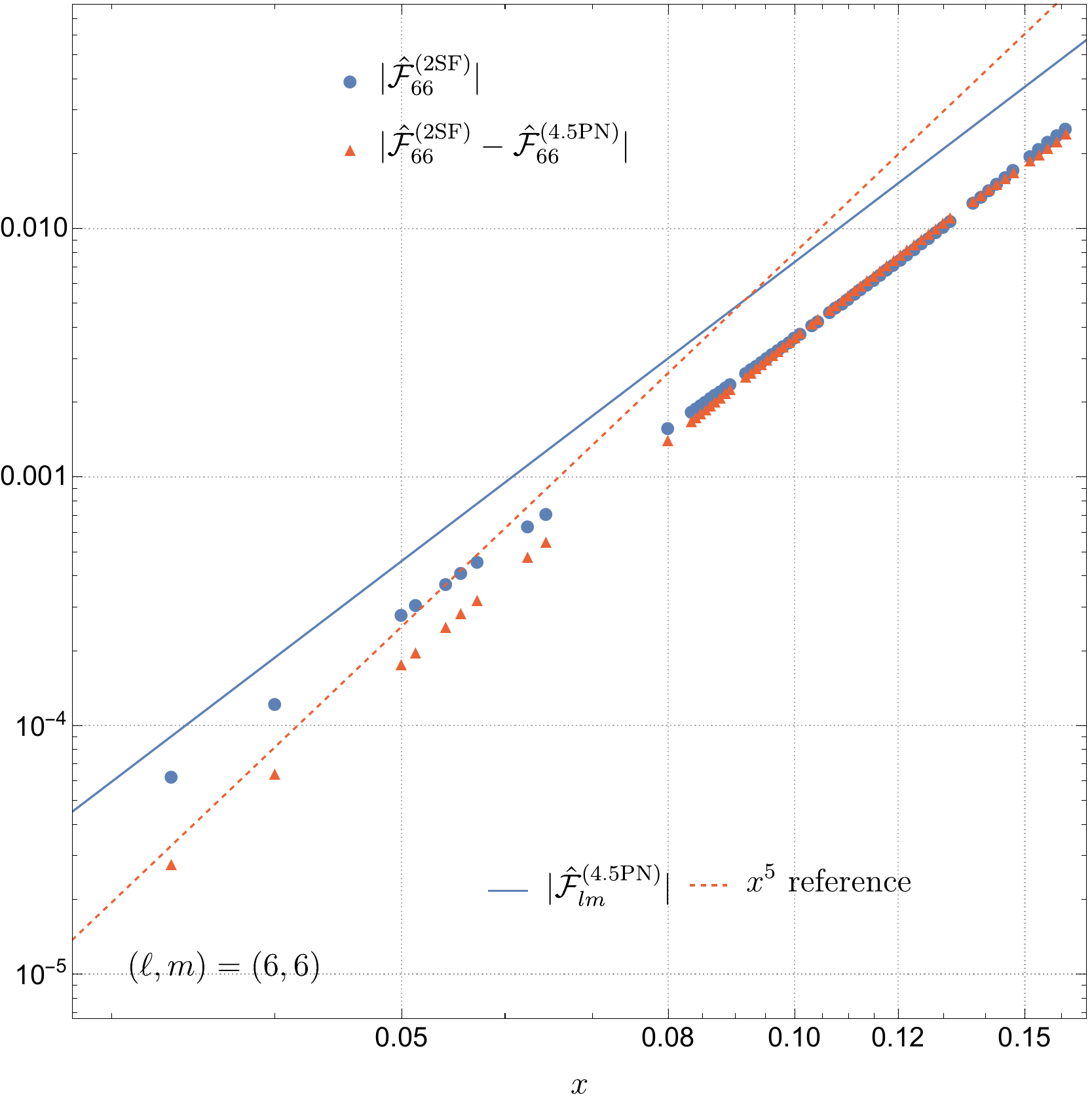} 
	\caption{Detailed comparison of $\hat{\mathcal{F}}^{(2)}_{55}$ and $\hat{\mathcal{F}}^{(2)}_{66}$, the $\mathcal{O}(\nu)$ coefficients of the $(5,5)$ and $(6,6)$ modes of the (Newtonian-normalized) flux, computed from (i) numerical GSF and (ii) PN theory.
    }\label{fig:comparisonmodes_l5m5_l6m6}
\end{figure*}

Comparisons for the individual modes can be found in: Fig.~\ref{fig:flux22} for the $(2,2)$ mode; Fig.~\ref{fig:comparisonmodes_l3m3_l3m2} for the $(3,3)$ and $(3,2)$ modes; Fig.~\ref{fig:fluxl4} for the $\ell=4$ modes; and Fig.~\ref{fig:comparisonmodes_l5m5_l6m6} for the $(5,5)$ and $(6,6)$ modes.
In contrast with the total flux, we find that we are able to resolve the expected scaling of the 4PN and 4.5PN residuals for many of these modes for small values of~$x$. However, one should note that the most challenging results to compute at 4.5PN (such at the renormalized source quadrupole moment, the tails of memory or the quartic tails) are entirely contained in the $(2,2)$ mode at this order. Moreover, we do not find clear agreement for all the $(\ell,m)$ modes, but as we highlight in Appendix \ref{apdx:1SFcomp}, there are sometimes zero crossings in the residuals that occur at very small values of $x$ and the correct residual scaling only becomes apparent for yet smaller values of $x$. 

We note here that the agreement between the $(\ell,m)$ modes of the GSF and PN calculations is not guaranteed, as
little work has been done to control the BMS frame in GSF calculations, which have mostly used intuition from the Newtonian-order analysis in Ref.~\cite{Detweiler:2003ci}. 
A BMS transformation, whether an ordinary Poincar\'e transformation or a supertranslation, can significantly alter the individual $(\ell, m)$ modes of the waveform~\cite{Boyle:2015nqa,Mitman:2024uss}, and this freedom remains unexplored in GSF calculations. 
Nonetheless, the good agreement we see at the level of modes suggests that the BMS frames of the PN and GSF calculations are not too distinct.
An analysis of the BMS frame of the GSF calculation is left for future work.

\section{Comparison at first-order in the mass ratio}\label{apdx:1SFcomp}

PN and GSF calculations will agree in the asymptotic regime of sufficiently small values of $x$. 
Exactly which values of $x$ are sufficiently small depends on the order at which the PN comparison is being made, and also the quantity, e.g., total flux or the flux for a given $(\ell,m)$ mode.
In particular, there can be zero crossings in the residual between the GSF data and the PN series such that the expected asymptotic behaviour in the residual will only be apparent for values of $x$ which are smaller than all the zero crossings.
These zero crossings can occur at quite small values of $x$, thus limiting our comparison with the 2SF data, as we only have accurate data from $x=0.02$ to $x\simeq0.162$.

\begin{figure*}
    \centering
    \includegraphics[width=0.48\linewidth]{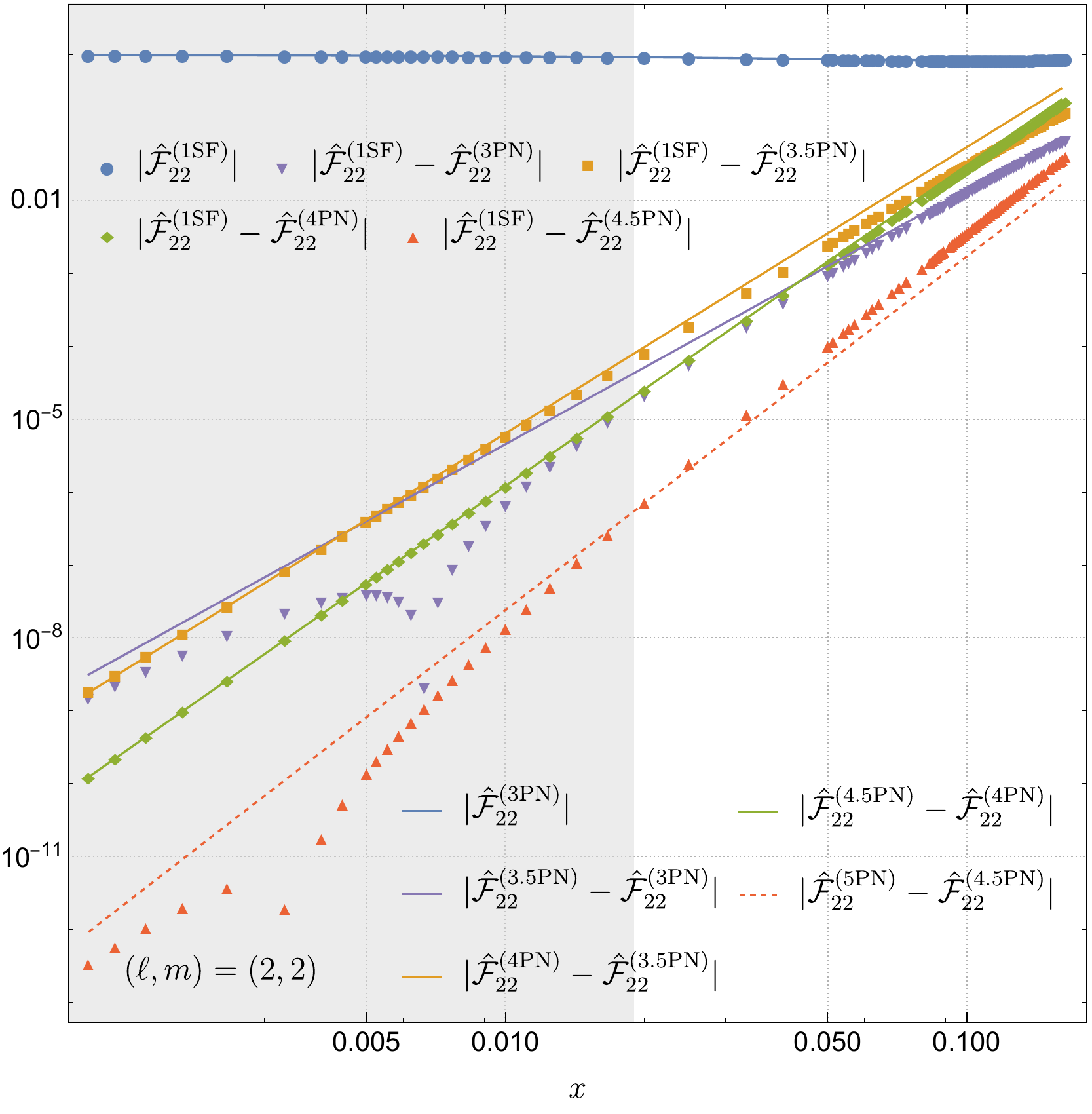}\hfill 
    \includegraphics[width=0.48\linewidth]{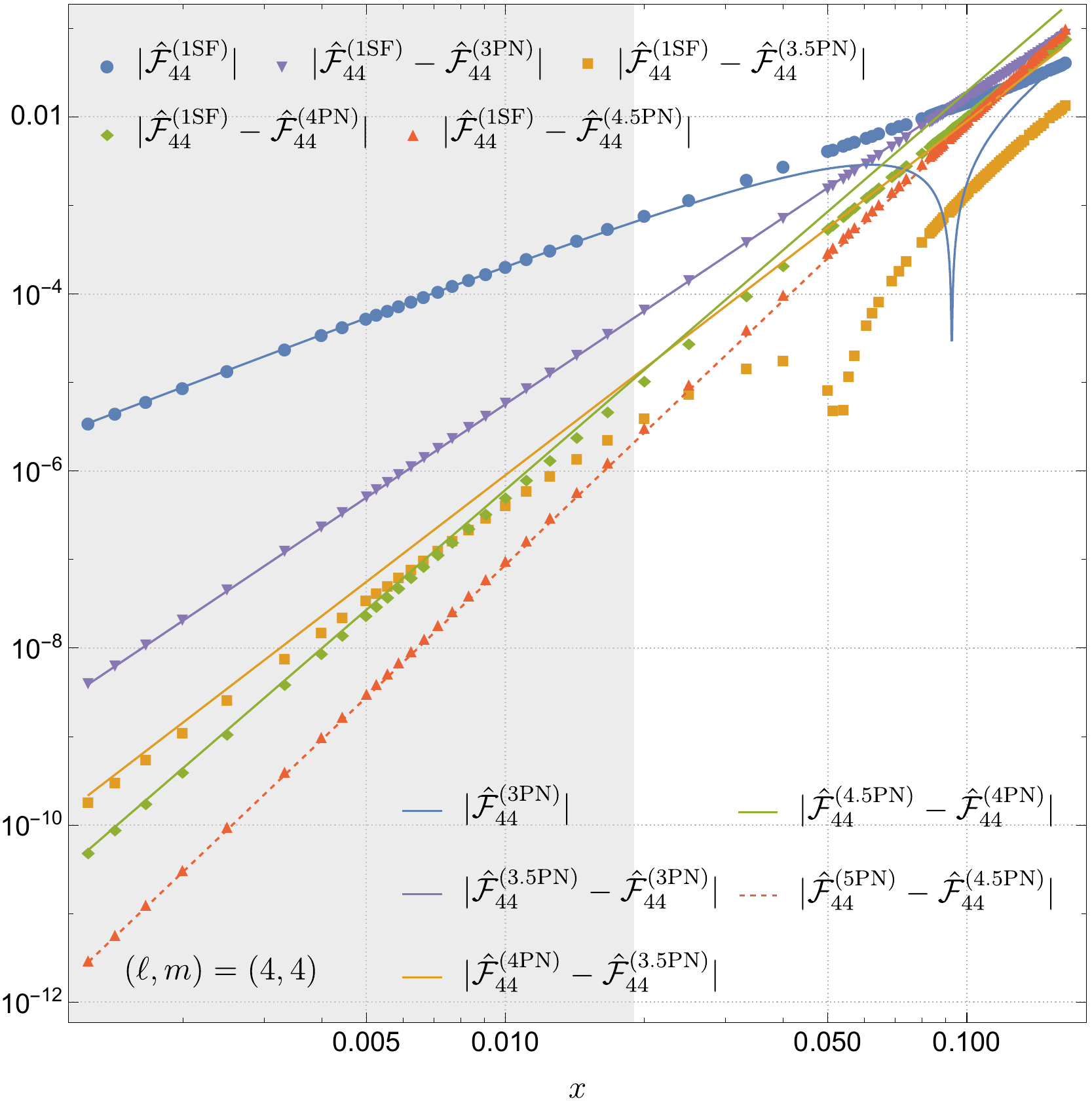}
    \caption{
    Detailed comparison of $\hat{\mathcal{F}}^{(1)}_{22}$ and $\hat{\mathcal{F}}^{(1)}_{44}$, the $\mathcal{O}(\nu^0)$ coefficients of the $(2,2)$ mode (left plot) and $(4,4)$ mode (right plot) of the (Newtonian-normalized) flux, computed from (i) numerical GSF and (ii) PN theory.
    The range of the 2SF data shown in the other figures in this paper is restricted to the white region between $x=0.02$ and $x\simeq0.162$.
    For the 1SF flux, we can accurately compute numerical data for $x < 0.02$, and this extra data is shown in the gray shaded region.
    For the $(2,2)$ mode, we see that the residuals with the 3PN and the 4.5PN series have zero crossings in the gray region at $x\simeq 0.0065$ and $x\simeq 0.0035$, respectively.
    The correct scaling of the residual only becomes apparent for yet smaller values of $x$. We also note that although there is a zero crossing in the residual with 3PN, further subtracting the 3.5PN and 4PN terms can result in residuals which settle more quickly to the expected scaling.
    For the (4,4) mode, we see that there is a zero crossing in the 3.5PN residual within the white region but that the correct scaling of the residual does not become apparent until very small values of $x$ (within the gray region). These two plots serve to highlight how zero crossings in the residual can delay the onset of the asymptotic regime where the PN and GSF results agree to very small values of $x$, and in particular smaller values of $x$ than we have 2GSF data for.
    }
    \label{fig:1SFvsPN}
\end{figure*}

To illustrate the effect of the zero crossings on the comparison, we take advantage of the fact that 1SF fluxes are known at much smaller value of $x$, and that the PN results at leading order in the mass-ratio have been also obtained at very high orders by GSF-PN methods \cite{TTS96}, see e.g. (A2)-(A27) therein. In Fig.~\ref{fig:1SFvsPN}, we show the comparison with the first-order flux for the (2,2) and (4,4) mode in Fig.~\ref{fig:1SFvsPN}. The numerical 1SF flux for these comparisons was computed using the \texttt{Teukolsky} package \cite{TeukolskyPackage} from the Black Hole Perturbation Toolkit \cite{BHPToolkit}.

\bibliography{reference_list.bib}
\end{document}